
\input harvmac

\def \inv {^{-1}}
\def \om {\omega}

\def\const {{\rm const}}
\def \s {\sigma}
\def\t {\tau}

\def \p {\phi}
\def \ha {\half}
\def \ov {\over}

\def \four{{\textstyle {1\ov 4}}}
\def \a {\alpha}
\def \lr {\lref }
\def\ep{\epsilon}

\def\vp {\varphi}
\def \bd {\bar \del}
\def \r {\rho}
\def\const {{\rm const}}\def\bd {\bar \del} \def\m{\mu}\def\n {\nu}\def\l
{\lambda}

\def \g {\gamma}

\def \y {{ \tilde y}}

 \def \sm {$\s$-model\ }
\def   \td {\tilde }
\def \k {\kappa}
\def \lr { \lref }

\gdef \jnl#1, #2, #3, 1#4#5#6{ { #1~}{ #2} (1#4#5#6) #3}

\def \M {{\cal M}}
\def \L {\Lambda}

 \def \h {\chi}
\def \da {\del_a}
\def \daa {\del^a}
\def \d {\del }

\def \y {{\td y}}
\def \b {\beta}
\def \S {{\cal S}}
\def \H  {{\cal H}}
\def \A {{\cal A}}
\def \cA {{\cal A}}

\def \cB {{\cal B}}

\lr \hrt {G.T. Horowitz and A.A. Tseytlin. \pr D50 (1994) 5204. }

\lr \roh { R. Rohm, \pr D32 (1985) 2849. }

\lr \cha{S. Chaudhuri and J.A. Schwarz, \pl B219 (1989) 291.}

\lr \cand{P. Candelas, G. Horowitz, A. Strominger and E. Witten, \np B258
(1985)  46.}
\lr \kounn {C. Kounnas, \pl B321 (1994) 26.}
\lr \gep {D. Gepner, \np B296 (1987) 757.}
\lr \kz{Y. Kazama and H. Suzuki,
\np B321 (1989) 232.}

\lr \gibma { G.W.  Gibbons and  K. Maeda, \np B298 (1988) 741.}
\lr\gib{G.W.  Gibbons, in: {\it Fields and Geometry}, Proceedings of the 22nd
Karpacz
Winter School of Theoretical Physics, ed. A. Jadczyk (World Scientific,
Singapore,  1986).}

\lr\gro{D.J.  Gross, J.A. Harvey,  E. Martinec and R. Rohm, \np B256 (1985)
253; \np B267 (1985) 75.}

\lr \green {M.B. Green, J.H.  Schwarz and E.  Witten, {\it Superstring Theory}
(Cambridge U.P., 1988).}

 \lr \kok {  E. Kiritsis and C. Kounnas,  ``Infrared behavior of closed
superstrings in strong magnetic and gravitational fields", hep-th/9508078. }

\lr \mme {M.A.  Melvin, \pl 8 (1964) 65. }
\lr \tser { A.A. Tseytlin, ``Exact solutions of closed string theory",
hep-th/9505052. }
\lr \gaun {F. Dowker, J.P. Gauntlett, G.W. Gibbons and G.T. Horowitz, ``The
decay of magnetic fields in Kaluza-Klein theory", hep-th/9507143.}
\lr\dowo{  F. Dowker, J.P. Gauntlett, D.A. Kastor and J. Traschen,
\pr D49 (1994) 2909; F. Dowker, J.P. Gauntlett, S.B. Giddings and G.T.
Horowitz, \pr D50 (1994)
2662.}

\lr \ruh {J.G. Russo and A.A. Tseytlin, ``Heterotic strings in a uniform
magnetic field", hep-th/9506071.}

\lr \kallosh { E.A.  Bergshoeff,  R. Kallosh and T. Ort\'\i n, \pr D47 (1993)
5444.  }
\lr \grs { M.B. Green and J.H. Schwarz,  B136 (1984) 307; \np B243 (1984) 285.
}

\lr \ghr { S. Gates, C. Hull and M. Ro\v cek, \np B248 (1984) 15.}
\lr \attick {J.J. Atick  and E. Witten, \np B310 (1988) 291. }

\lr  \rut{J.G. Russo and A.A. Tseytlin,  \np B449 (1995) 91, hep-th/9502038.}
\lr \melvint{A.A. Tseytlin,  \pl B346 (1995) 55. }
\lr \rts{J.G. Russo and A.A. Tseytlin, \np B448 (1995) 293,
hep-th/9411099.}

 \lr \gross {D.J.  Gross, J.A. Harvey,  E. Martinec and R. Rohm, \np B256
(1985)
253; \np B267 (1985) 75.}

\lr \gso{
M.B. Green, J.H.  Schwarz and E. Witten, {\it Superstring Theory} (Cambridge
U.P., 1987). }

\lr \ssu {E. \' Alvarez, L. \'Alvarez-Gaum\'e and I. Bakas,
``T-duality and space-time supersymmetry", hep-th/9507112;
K. Sfetsos, ``Duality and restoration of manifest supersymmetry",
hep-th/9510034.
}

\lr\kal{I. Bakas, \pl B343 (1995) 103; E. Bergshoeff, R. Kallosh and T.
Ort\'in, \pr D51 (1995) 3009.}
\lr\bak{I. Bakas and K. Sfetsos, \pl B349 (1995) 448;
S.F. Hassan, ``T-duality and non-local supersymmetries", hep-th/9504148.}

\lr \sen{A. Sen, \pr D32 (1985) 2102; \prl 55 (1985) 1846.}
\lr \hulw { C. M. Hull and E.  Witten,  \jnl \pl, B160, 398, 1985. }
\lr \bur{C.P. Burgess, \np B294 (1987) 427; V.V. Nesterenko, \ijmp A4 (1989)
2627.}

\def \np {  Nucl. Phys. }
\def \pl { Phys. Lett. }
\def \mpl { Mod. Phys. Lett. }
\def \prl { Phys. Rev. Lett. }
\def \pr  { Phys. Rev. }

\def \ijmp { Int. J. Mod. Phys. }

\baselineskip8pt
\Title{\vbox
{\baselineskip 6pt{\hbox{  }}{\hbox
{Imperial/TP/94-95/62 }}{\hbox{hep-th/9510041}} {\hbox{
   }}} }
{\vbox{\centerline {Closed superstrings in magnetic field:}
 \centerline {instabilities and supersymmetry breaking  }
}}

\vskip -20 true pt

\medskip
\centerline{   A.A. Tseytlin\footnote{$^{\dagger}$}{\baselineskip8pt
e-mail address: tseytlin@ic.ac.uk. \ On leave  from Lebedev  Physics
Institute, Moscow.} }

\smallskip\smallskip
\centerline {\it  Theoretical Physics Group, Blackett Laboratory,}
\smallskip

\centerline {\it  Imperial College,  London SW7 2BZ, U.K. }
\bigskip
\centerline {\bf Abstract}
\medskip
\baselineskip10pt
\noindent
We consider a 2-parameter class of solvable closed superstring models  which
`interpolate' between Kaluza-Klein and dilatonic
Melvin magnetic flux tube backgrounds. The spectrum of string
states has similarities with  Landau spectrum
for a charged particle in a uniform magnetic field.
The presence of spin-dependent `gyromagnetic' interaction
implies breaking of supersymmetry and possible existence
(for certain values of magnetic parameters) of tachyonic instabilities.
We study in detail the simplest example of the Kaluza-Klein Melvin model
describing a superstring moving in flat but non-trivial
10-d space containing a 3-d factor which is a `twisted' product
of a 2-plane and an internal circle. We also discuss the compact
version of this model constructed by `twisting'
the product of the two groups in $SU(2) x U(1)$ WZNW theory
without changing the local geometry (and thus the central charge).
We explain how the supersymmetry is broken by continuous
`magnetic' twist parameters and comment on possible implications
for internal space compactification models.

\bigskip
\bigskip
\bigskip

\centerline{ To appear in: {\it ``String Gravity and Physics at the Planck
Scale",}}
\centerline{ Proceedings of  Chalonge School, Erice, 8-19 September 1995,  ed.
N. S\'anchez (Kluwer, Dordrecht)}

\medskip

\Date {October 1995}

\noblackbox
\baselineskip 14pt plus 2pt minus 2pt
\vfill\eject

\def \N {{\hat N}}
\def \X {{\cal X}}

\lr \witte{ E. Witten, \np B266 (1986) 245;
M.T. Grisaru, P. Howe, L. Mezincescu, B.~Nilsson and P.K.  Townsend,
\pl B162 (1985) 116;
J.J. Atick, A. Dhar and B.~Ratra, \pl B169 (1986) 54. }
\lr \fra{E.S. Fradkin and A.A. Tseytlin, \pl B158 (1985) 316;
\pl B160 (1985) 69.}

\lr\bach{ C. Bachas, ``A way to break supersymmetry", hep-th/9503030.}

\lr \kiri {E. Kiritsis and  C. Kounnas, \np  B442 (1995) 472;
 ``Infrared regulated string theory and loop corrections to coupling
constants", hep-th/9507051. }

\lr \alvi{L. Alvarez-Gaum\' e, G. Moore and C. Vafa, {Commun. Math. Phys.} 106
(1986) 1. }

\lr \rohm {R. Rohm, \np B237 (1984) 553.}
\lr \bri {M.B. Green, J.H. Schwarz and L. Brink, \np B198 (1982) 474.}

\lr \giva{P. Ginsparg and C. Vafa, \np {B289 }  (1987) 414.}

\lr\sche{ J. Scherk and J.H. Schwarz,
 Phys. Lett. { B82} (1979) 60; Nucl. Phys. {B153} (1979) 61.}
\lr\sch{S. Ferrara, C. Kounnas and M. Porrati, \np B304 (1988) 500; Phys. Lett.
{ B206} (1988) 25;
C. Kounnas and M. Porrati, Nucl. Phys. { B310} (1988) 355.}

\lr \koun {V.P. Nair, A. Shapere, A. Strominger and F. Wilczek, \np B287 (1987)
402; P. Ginsparg and C. Vafa, \np B289 (1987) 414; H. Itoyama and T.R. Taylor,
Phys. Lett. { B186} (1987) 129.}

\lr \att {J.J. Atick and E. Witten, \np B310 (1988) 291.}
\lr \kou {C. Kounnas and B. Rostant, \np B341 (1990) 641.}
\lr\diens {K. R. Dienes, \np B429 (1994) 533; hep-th/9409114;
hep-th/9505194.}

\lr\rrt{J.G. Russo and A.A. Tseytlin, ``Magnetic flux tube models in
superstring theory", hep-th/9508068.}

\lr \abou{ A. Abouelsaood, C. Callan, C. Nappi and S. Yost, \np {B280 }
 (1987) 599.}

\lr \frat{ E.S. Fradkin and A.A. Tseytlin, \pl {B163  }
(1985) 123.}

\lr \fer{ S. Ferrara and M. Porrati, \mpl {A8 }
 (1993) 2497.}

\lr \niel{ N.K. Nielsen and P. Olesen,
\np {B144 }
 (1978) 376. }

\lr \ruu{ J.G. Russo and L. Susskind,   \np {B437  }
(1995) 611;  A. Sen, \np {B440} (1995) 421. }

\lr \tomb{M. Porrati and E.T. Tomboulis, \np B315 (1989) 615;
E.T. Tomboulis, \pl B198 (1987) 165.}

\lr \anton { I. Antoniadis, C. Bachas, D. Lewellen and T. Tomaras, \pl B207
(1988) 441.}
\lr \band{T. Banks and L. Dixon, \np B307 (1988) 93.}

\lr \barny{I. Bars, D. Nemeschansky and S. Yankielowicz, \np B278 (1986) 632. }

\lr \abr {
H.B. Nielsen and P. Olesen, \np { B61 }
 (1973) 45.  }

\lr \aaan{I. Antoniadis, \pl B246 (1990) 377;
I. Antoniadis, C. Munoz and M. Quiros, \np B397 (1993) 515;
K. Benakli, ``Perturbative supersymmetry breaking in orbifolds with Wilson line
backgrounds", hep-th/9509115.}

\lr \deal{S. de Alwis, J. Polchinski and R. Schimmrigk, \pl B218 (1989) 449.}

\lr \din {M. Dine and N. Seiberg, \np B301 (1988) 357.}

\lr \roh { R. Rohm, \pr D32 (1985) 2849. }

\lr \cha {S. Chaudhuri and J.A. Schwarz, \pl B219 (1989) 291.}
\lr\kal{E. Bergshoeff, R. Kallosh and T. Ort\'in, \pr D51 (1995) 3009;
I. Bakas, \pl B343 (1995) 103.}
\lr\bak{I. Bakas and K. Sfetsos, \pl B349 (1995) 448.}
\lr \hulh{S.J. Gates, C.M. Hull and M. Ro\v cek, \np B248 (1984) 157;
P. Howe and G. Sierra, \pl B148 (1984) 451.}

\lr \hass {S.F. Hassan and A. Sen, \np B405 (1993) 143;
M. Henningson and C. Nappi, \pr D48 (1993) 861;
E. Kiritsis, \np B405 (1993) 109.}

\newsec{Introduction}
An important problem in string theory is to study
how quantised strings propagate  in non-trivial backgrounds
described by conformal 2d models.
This may help to understand the structure of the space of exact string
solutions as well as certain generic properties of
string theory like the existence `critical' or  `maximal'
values of fields,  instabilities,
mechanisms of supersymmetry breaking, residual symmetries of string spectrum,
etc.

Magnetic field  is  one of the simplest
probes of the spectrum and critical properties of a  physical system.
In the context of a  gravitational theory like closed string theory,
a magnetic background is accompanied by a curvature of space-time.
In spite of that,
 certain  closed string magnetic models can be solved exactly.  In particular,
these  are static flux tube
type configurations  with approximately uniform magnetic field   which
generalize
the Melvin solution of the Einstein-Maxwell theory
\mme.
Such backgrounds are exact solutions of (super)string theory
\refs{\melvint,\rut}
and, moreover, the spectrum of the corresponding  unitary
conformal string models  can
be explicitly determined \refs{\rut,\rrt}.

Below we shall first review  the solution \rrt\ of this  class of magnetic
models.  In the last section we shall discuss   related
compact models which may be used for string compactifications
and
 the issue of supersymmetry breaking induced by
`magnetic' twist parameters.

Before turning to closed string theory let us first recall the solution
of similar uniform  magnetic field problems in particle (field) theory and open
string theory.

\subsec{Point particles  and open strings  in  magnetic field}
The reason why the  quantum-mechanical or field-theoretical problem of  a
particle in  a  uniform abelian
(electro)magnetic field is exactly solvable
 is that the action  $I= \int d\t  [\dot x^\m \dot x^\m   +i
e \dot x^\m A_\m (x)]$  (which determines the Hamiltonian in quantum mechanics
and the  heat kernel in field theory)
becomes gaussian if the field  strength is constant,
 $A_\m =-\ha F_{\m\n} x^\n,   \ \ F_{\m\n} = const  . $
Assuming the magnetic field is directed along $x^3$-axis
(so that $F_{ij} = H \ep_{ij}$, $i,j=1,2$)
and introducing $x=x_1 + ix_2$,\  $a=-i(\del_x^* + \four  eH x), \
a^\dagger=-i(\del_x  -  \four  eH x^*), $\
$[a, a^\dagger] = \ha eH$, one can reduce the corresponding quantum-mechanical
problem to a free oscillator one. The resulting  energy spectrum is
the special ($S=0$) case   of the Landau spectrum for a particle of charge $e$,
mass $M_0$ and  third  component of the spin $S$ (we assume $eH > 0$)
\eqn\ene{  E^2 = M^2_0 +  p_3^2    - 2   e { H} J \ ,  \ \ \  J\equiv  - l -
\ha  + \ha g S \ ,  \ \ \ \  l=0,1,2, ...\ . }
Here $l$ is the Landau level number (which replaces the continuous momenta
$p_1,p_2$) and  $g$ is the
gyromagnetic ratio  which is  $1/S$ for minimally coupled particles but can be
equal to 2 for non-minimally coupled ones.
Thus $E^2$ can become negative for large enough values of $H$,
e.g., $H > H_{crit}= M^2_0/e$ for spin 1 charged states.
That applies, for example, to  $W$-bosons in the context of electroweak theory
\niel\
suggesting the presence of a  transition to a phase with a
$W$-condensate.
In the case of unbroken gauge theory with massless
charged vector particles  the instability is present for any (e.g.,
infinitesimal) value of the magnetic field \niel.
This
infra-red  instability of a magnetic background is not cured
 by supersymmetry, i.e.
it remains also in supersymmetric gauge theories
(e.g.,  in ultra-violet finite $N=4$ supersymmetric
  Yang-Mills theory)
since the small fluctuation operator for the gauge field $-\delta_{\m\n} D^2 -
2 F_{\m\n} $
still has negative modes
  due to the `anomalous magnetic moment' term.
This is not surprising  given that
 the magnetic spin-dependent coupling  breaks
supersymmetry.

The meaning of this instability is that the expansion
is carried out  near a classical solution of the Yang-Mills equations
(abelian $A_i = - \ha H \ep_{ij} x^j$  belonging to the  Cartan subalgebra)
which is {\it not } a vacuum one:
the energy is proportional to $H^2$ and thus is minimal for $H=0$.
As a result, a non-zero magnetic field will tend to dissipate.
The  presence of  tachyonic modes in the magnetic background with
infinitesimal (but non-vanishing $H$) does not  indicate an instability
of the trivial vacuum but only that of a configuration with $H\not =0$ (this
corresponds to a resummation of the expansion  in small perturbations
near   the
trivial vacuum).

The solution of the uniform magnetic field problem in the open string theory is
similar to the one in the particle case. Indeed, the open  string is coupled to
the  (abelian) vector field only at its ends, $
I=  {1\ov 4\pi \a'} \int d^2 \s \  \del_a x^\m \del^a x^\m  +  ie \int d\tau \
A_\m (x) \dot x^\m ,  $
 and
thus  for the abelian  background
$A_\m =-\ha F_{\m\n} x^\n$  the resulting   gaussian path integral can be
computed exactly \frat.
This is a consistent `on-shell' problem  since $F_{\m\n}=\const$ is
an exact  solution of the effective field
equations  of the open string theory.
The  corresponding 2d world-sheet theory  is a conformal field theory \abou\
which can be solved
explicitly in terms of free oscillators (thus representing  a
string generalization  of the   Landau  problem   in quantum particle
mechanics).
As a result, one is able to  determine the spectrum of an open  string moving
in a constant magnetic field \refs{\abou,\bur,\fer}.
 The   Hamiltonian  ($L_0$) of the
open  (super)string  in a constant magnetic field is given by \refs{\abou,\fer}
\eqn\ope{
\hat \H =
 \ha  \a' \big( -E^2 +  p_\a^2) +\hat N -    \g \hat J
\ , }
$$   \hat \g \equiv {2 \ov \pi} \big|{ \rm arctan} (\ha \a' \pi e_1 H) +
{\rm arctan} (\ha \a' \pi e_2  H )\big| \ , \ \ \  \  0 \leq  \hat \g < 1 \ .
$$
Here $e_1,e_2$ are charges at the two ends of the  string
and $H$ is
 the magnetic field, $F_{ij}= H \epsilon_{ij}$.
$\hat N=0,1,2...$ is the number of states operator  and
$\hat J = - l - \ha  + S $  is the angular momentum operator
of the open string in the $(1,2)$ plane.
The  energy spectrum  is found from the constraint $\hat \H=0$ and
 for small $e_i H$
is in agreement with   \ene\ with $g=2$.
A novel  feature of this spectrum as
 compared to the free  (super)string  one
 is the presence of  tachyonic states  above certain critical values of the
magnetic field ($H_{crit}\sim \a'^{-1}   \hat N/S \sim \a'^{-1}$).  Thus
 the constant magnetic field background  is unstable
 in the open string theory (as it  is   in the  non-abelian
gauge theory).  A qualitative  reason for this instability is that the
 free  open string  spectrum
contains  electrically  charged higher spin
massive particle states and the latter    have  (approximately)
 the Landau spectrum \ene\ ($\hat \g \approx \a' (e_1 + e_2)H$ for a weak
field).
The tachyonic states appear only in the bosonic
   (Neveu-Schwarz)  sector  and only on  the leading
Regge trajectory ($l=0$, maximal $S$ for a given mass level $\hat N$,
$\  S= \hat N + 1$).

\subsec{Closed strings and Melvin-type magnetic flux tube backgrounds}

Similar background magnetic field problem is  also exactly solvable
in  {\it closed}  (super)string theories \refs{\rts,\rut,\rrt}.
This may  be  unexpected  since
  the abelian vector  field  is now
coupled to the  internal points of the string,
and such interaction terms, e.g.,
$\Delta L =  \del y \bd y +  A_\m (x) \del x^\m \bd y +
\td A_\m (x) \bd x^\m \del y + ... , $
do not become gaussian  for $A_\m =-\ha F_{\m\n} x^\n$.
Here $y\in (0, 2\pi R)$ is a compact internal Kaluza-Klein coordinate
that `charges'  the string. The  spectrum of the closed string compactified on
a circle
contains  states with arbitrarily large masses, spins
{\it  and } charges $Q_{L,R} = mR\inv  \pm w\a'^{-1} R , \ (m,w =0, \pm 1, ...)
$.
One difference  compared to
the open string case is that there  are infinitely many possible values of
charges. Another is that
both $Q_L$ and $Q_R$ are,  in general,
 non-vanishing  and coupled to (a combination of)
two
abelian vector fields ($G_{5\m}$ and $B_{5\m}$).\foot{As a result, the
gyromagnetic ratio
of closed string states is $g \leq 2$, e.g., $g=1$ for standard non-winding
Kaluza-Klein states \refs{\ruu,\rut}.}

The important observation is  that in contrast to the  tree level
abelian
open string case, the
$F_{\m\n}=\const$ background in flat space does not
represent a solution of  a  {\it closed} string theory,
i.e. the above interaction terms  added to the  free string Lagrangians
do not lead to   conformally invariant 2d  $\s$-models.
Since the closed string  theory contains gravity,
 a uniform  magnetic field,  which has a non-vanishing energy,
must curve the space (as well as
possibly induce other `massless' background fields).
One should thus first find a consistent  conformal model
which is a closed string analogue
of the    uniform magnetic field background  in  the
 flat space  field (or open string) theory
and  then address the question of its solvability.
Remarkably, it turns out that extra terms  which
should be added  to   the  closed string action in order to  satisfy
 the conformal invariance condition (i.e. to satisfy the closed string
effective field  equations)
produce  solvable  2d models.
Like  the particle and open string  models,  these  models are
exactly solvable  in terms of free oscillators (one is able to find explicitly
the spectrum, partition function, etc).

A  simple example of an approximately   uniform
  magnetic  field background
  in  the Einstein-Maxwell
theory is   the static cylindrically symmetric
 Melvin `magnetic universe' or `magnetic flux tube'  solution \mme.
It has $R^4$  topology  and   can be  considered \gib\    as
a gravitational analogue of the  abelian Higgs model vortex
\abr\
with the magnetic pressure (due to  repulsion of Faraday's flux lines)
being balanced not by the  Higgs field but by  the gravitational attraction.
The magnetic field is  approximately constant inside  the
tube and decays  to zero  at infinity in the direction orthogonal to
$x_3$-axis. The  space is not, however,
asymptotically euclidean: the $(\r, \vp)$   2-plane orthogonal
to $x_3$  asymptotically closes at large $\r$ (so the solution should be
interpreted not as a flux tube embedded in  an approximately flat space,
but rather as a  `magnetic universe').
 Several interesting  features  of the Melvin  solution in the context of
Kaluza-Klein (super)gravity (e.g.,  instability against monopole  or  magnetic
black hole pair creation)  were discussed  in \refs{\gib,\dowo,\gaun}.

This  Einstein-Maxwell (`$a=0$') solution
has  two  direct   analogues     among solutions of  the
 low-energy closed string  theory (superstring toroidally
compactified to $D=4$).
Assuming $x^5=y$ is a compact internal coordinate,
 the   $D=5$ string effective
action  can be expressed in terms of $D=4$ fields:
 metric $G_{\m\n}$, dilaton $\p$,  antisymmetric tensor
 $B_{\m\n}$, {  two}  vector fields ${\cal A}_\m$ and ${\cal B}_\m$ (related to
$G_{5\m} $ and $B_{5\m}$)   and
the `modulus'  $\s$. The  dilatonic (`$a=1$')  and  Kaluza-Klein (`$a=\sqrt
3$')
Melvin solutions have zero
$B_{\m\n}$  but  $\p$ or $\s$ being non-constant.
 Starting
 with the $D=5$ bosonic string effective action and
 assuming  that one spatial dimension $x^5$ is compactified on a small
circle,  one
finds (ignoring massive Kaluza-Klein modes)
the following dimensionally reduced  $D=4$  action
\eqn\actt{ S_4 = \int d^4 x \sqrt { G }\  e^{-2\Phi}    \ \big[
  \   R \ + 4 (\del_\m \Phi )^2  - (\del_\m \s )^2  }
 $$- \ {\textstyle {1\over 12}} (\hat H_{\m\n\l})^2\  -  \four e^{2\s} ({
F}_{\m\n}
({\cal A}))^2
-\four e^{- 2\s} (F _{\m\n} ({\cal B}))^2
  + O(\a')   \big]  \  , $$
 where $\hat G_{\m\n}= G_{\m\n} - G_{55} \A_\m \A_\n, $ \ $ F_{\m\n} ({\cal A})
= 2\del_{[\m} {\cal A} _{\n]}, $    $ \ F_{\m\n}({\cal B}) = 2 \del_{[\m} {\cal
B}_{\n]}, $
$\   \hat H_{\l\m\n} = 3\del_{[\l} B_{\m\n]} - 3 {\cal A}_{[\l} F_{\m\n]}
({\cal B})$,  $ \  \Phi= \phi - \ha \s .$
 The effective  equations following from  \actt\
 have a
3-parameter $(\a,\b,q)$ class of stationary axisymmetric
(electro)magnetic flux tube solutions \rut.
The most interesting   subclass   of  these  solutions ($\a=\beta$) describes
 {\it static}  magnetic flux tube  backgrounds ($\b$ and $q$ are magnetic field
strength parameters). It  contains
the $a=\sqrt 3$  and $a=1$  Melvin solutions  \gibma \  as the special cases of
$\beta=0$   and $\beta=q$. The four-dimensional geometry is given by (in  terms
of the string
metric in \actt; $\ x^\m=(t, \r, \vp, x^3)$)
\eqn\backg
{ ds^2_4=  - dt^2  + d\r^2 +   F(\r)\td F(\r)  \r^2
d\vp^2 + dx_3^2 \ ,\
}
\eqn\baag{ {\cal A}_\vp=  q  F(\r)   \r^2 \ , \ \ \ \
{\cal B}_\vp=  -\beta \td F(\r)   \r^2 \ , \ }
$$
e^{2(\phi-\phi_0)}=\td F(\r )\ ,\ \   e^{2\s} = \td F(\r) { F}\inv
(\r) \ , \ \ \  F  \equiv {1\ov 1 +  q^2 \r^2}  \  ,\ \ \
\td F  \equiv {1\ov 1 + \beta ^2 \r^2} \ .
$$
There are two magnetic fields with  the effective strengths
$q$ and $\b$ (when both are non-vanishing the radius of $\vp$-circle
goes to zero at large $\r$).
In particular,\foot{In these special cases
there is effectively  just one non-trivial vector and one scalar
so that the Einstein-frame action can be put into the
form $\int d^4 x \sqrt { G' }\  \big[   R'\ - \ha (\del_\m \psi  )^2
  -  \four e^{-a\psi } {  F}_{\m\n}^2 \big]  .$}
the `$a=\sqrt 3 $ Melvin' background is   ($\b=0, \ q\not=0$):
\eqn\back{  \ ds^2_4=  - dt^2  + d\r^2 +   F(\r)  \r^2 d\vp^2 + dx_3^2 \ ,  }
$$ {\cal A}=  q  F(\r)   \r^2 d\vp \ ,
\  \ \ e^{2\s} = { F}\inv = 1 +  q^2 \r^2 \ , \ \ \  \ {\cal B}= B=0\ , \ \ \
 \p=\p_0  \ ,      $$
and the  `$a=1$ Melvin' background is   ($\b=q\not=0$):
\eqn\meq{ ds^2_4=  - dt^2  + d\r^2 + \td F^2(\r)  \r^2 d\vp^2 + dx_3^2 \ ,  }
$$ {\cal A}= -{\cal B} =  \b \td F(\r)   \r^2 d\vp\ , \
\ \ \  e^{2(\p-\p_0)}=  \td F  = (1 + \b^2 \r^2)\inv  \  ,  \ \  \ \ B= \s=0\ .
    $$
The  string model
corresponding to the $a=1$  Melvin background \gibma\  was  constructed  in
\melvint\ and solved in \rut.  The general case of arbitrary  $(\b,q)$
was studied in \rut.
The model describing $a=\sqrt 3 $ Melvin background
is the simplest possible special case.
The reason is that when viewed from higher   dimensions
the background \back\
 corresponds to a {\it flat}  (but globally non-trivial)
5-dimensional space-time \refs{\dowo,\gaun}.\foot{This is an
interesting example of a Kaluza-Klein  background which looks
non-trivial (curved) from lower-dimensional point of view but is actually flat
as 5-dimensional space. That means that the total contribution of the  three
4-dimensional fields (metric, vector and scalar)
in any local observable will vanish. However, the global properties
of the corresponding string theory will be non-trivial. Note also
that since the effective Kaluza-Klein radius $e^\s R$ grows with
$\r$,  the Kaluza-Klein interpretation may not apply for large
$q$ (see  \gaun).}
This is why the associated string model
is explicitly solvable.

The remarkable simplicity of the  $a=\sqrt 3 $ Melvin string model  makes it  a
good pedagogical example
which we shall discuss first (Section 2), before turning
to more general
$(\b,q)$ models (which no longer correspond to flat higher dimensional spaces
but  are still solvable).
 The  quantum Hamiltonian
 of the corresponding type II superstring model will be equal to
the free superstring one plus terms linear and quadratic in
 angular momentum operators.
As a result, the mass spectrum can be explicitly determined.
We  will show that supersymmetry is   broken and that there  exist
intervals of values of moduli parameters
(Kaluza-Klein radius and magnetic field strength)
for which the model is unstable. The string
partition function on the torus is IR finite  or infinite depending on
the values of the parameters.
We shall consider  both the Ramond-Neveu-Schwarz and the light-cone
Green-Schwarz
 formulation of the
theory (in the
latter the  breaking  of supersymmetry is related to the
absence of Killing spinors in the   Melvin background).

In Section 3 the results
 obtained for the $a=\sqrt 3$ Melvin model
will be generalized to the  $(\b,q)$ class of static magnetic flux tube  models
and, in particular,  to the $a=1$ Melvin model.
We shall explain the reason for solvability of these models
and clarify the nature of perturbative
instabilities that appear for generic values  of the magnetic
field  parameters.

In Section 4 we shall
first  explain   the
 relation  between  the $a=\sqrt {3}$ Melvin  model
and   superstring compactifications on twisted tori
where supersymmetry is broken by discrete rotation
 angles.  Then we shall
consider the  compact
version of the $a=\sqrt 3 $ Melvin model
which is obtained by `twisting'  $SU(2) \times U(1)$ WZNW model
and discuss  the issue of supersymmetry breaking
by  the twist parameters.

\newsec{Superstring  model  for  $a=\sqrt 3$ Melvin background }


\subsec{Bosonic string model}
Let us consider the closed bosonic string propagating in
the space $M^D= M^5 \times T^{D-5}$, where $T^n$ is a torus
and $M^5= R_t \times R_{x_3} \times \M^3$. \
$\M^3$ is flat but globally non-euclidean space
which can be represented as a twisted product
(symbolically, $\M^3= R^2 \ast S^1$)   of the  2-plane  $R^2$ $(\r,\vp)$
 with Kaluza-Klein circle  $S^1$ $(y \in (0, 2\pi R))$
(or as a  bundle with $S^1$ as a base and $R^2$ as a fibre).
 It can also be  obtained \gaun\
by factorizing $R^3$ over the group generated by
 translations in  two angular directions:
in the coordinates where $ds^2=
d\r^2 + \r^2 d\theta^2 + dy^2 $ one should
identify the points $(\r,\theta, y) = (\r, \theta + 2\pi n   + 2\pi qRm, y+
2\pi R m)$ \ ($n,m=$integers), i.e. combine the shift  by $2\pi R$ in $y$ with
a rotation by an arbitrary angle $2\pi qR$ in the 2-plane.
In terms of the globally defined $2\pi$-periodic coordinate $\vp $
the metric of $\M^3$ is
\eqn\metr{ds^2 = d\r^2 + \r^2 (d\vp +q dy)^2 + dy^2 \ .
 }
It  is  flat  since  locally one may
introduce the coordinate $\theta= \vp + q y$ and decouple $y$ from
$\r, \vp$.
The global structure of this 3-space is non-trivial:
the fixed $\r$ section is a 2-torus (with $\r$-dependent conformal factor and
complex modulus) which degenerates into a circle at $\r=0$ (the
space  is  actually regular everywhere, including  $\r=0$).

The Lagrangian describing string propagation in such  flat but globally
non-trivial  $M^D$ is
\eqn\mode{ L=L_0 + L_1 \ , \ \ \ \ \ \
L_0 = -\d_a t \d^a t + \d_a x_\a \d^a x^\a \ , \ \ \ \ }
\eqn\mee{
L_1 (\M^3)  = \d_a \r \d^a \r + \r^2 (\d_a \vp + q \da y)(\daa \vp + q \daa y )
 + \da
y \daa y  \ . }
Here $\r\geq 0 $ and $0 < \vp \leq 2\pi $   correspond to the cylindrical
coordinates  on a ($x_1,x_2)$-plane, $y$ is a  `Kaluza-Klein'
coordinate
with period $2\pi R$, and $x^\a $ include the  flat $x^3$-coordinate
of $D=4$ space-time and, e.g., 21  (or 5 in  the superstring case)  internal
coordinates compactified on a torus.
To give the
4-dimensional  interpretation to this model  $L_1$ should be  represented in
the  `Kaluza-Klein'
form
 \eqn\me{
L_1 = \d_a \r \d^a \r +  F(\r) \r^2 \d_a \vp \daa \vp
  +  e^{2\s}  ( \da y   +  {\cal A}_\vp \da \vp)   ( \daa y   +  {\cal A_\vp}
\daa \vp)    \ ,  }
where  $F\inv= e^{2\s} = 1 + q^2 \r^2 , \ \A_\vp = q\r^2 F$.
The resulting  $D=4$ background (metric, Abelian vector  field $\cal A_\m$ and
scalar $\s$)  is indeed the $a=\sqrt 3$ Melvin geometry \back.
The parameter $q$ thus has the interpretation of the
magnetic field strength
at the core of the flux tube. From the 4-dimensional point of view this model
describes the motion
of charged string states in the  $a=\sqrt 3$ Melvin  magnetic flux tube
background.\foot{We  assume that $q$ is a continuous
parameter. There is no reason for its quantization at the level of string
model \mee.  Given the magnetic flux tube  \back\ interpretation of this model,
one may, however, wonder how this is reconciled
with  the flux quantization in similar magnetic backgrounds
like the  Higgs scalar vortex or magnetic monopole on 2-sphere.
Though the magnetic flux through the $(\r,\vp)$ 2-plane
is finite in the case of the Melvin background,
the topological argument for its quantization (cf. \gib)
does not apply since the  2-space is non-compact (the $\r=\infty$ point
is not part of the space).
Given a  magnetic field configuration
with a finite flux through a 2-plane,
the flux may be  quantised  once charges are
added  since this may corresponds to   a state of minimal energy
(cf. the case of the Cooper pairs in superconductors;
the minimal energy condition leads  also to the  asymptotic
condition $D_\m\phi =0, \  ieA_\m=\del_\m \phi $, etc.,  in the case of the
scalar vortex).}

Although the 5-space is flat, the   string theory  \mee\ will  be
non-trivial
already at the classical level (due to the existence of winding string states)
and  also at the quantum level
in the non-winding sector (where there will be  a `magnetic' coupling
to the  total angular momentum in the 2-plane). This
represents  an   example of a gravitational
5d (space-time)  Aharonov-Bohm-type  phenomenon: the value of the magnetic
field strength parameter  $q$ does not
influence the (zero) curvature of the space but affects the  global properties
like masses of string states.

Since $\M^3$ is flat, the model \mee\ is conformal for arbitrary
values of the two parameters  $q$ and $R$.
Certain values of these  moduli are special:
if $ qR = n, \  \ n= 0, \pm 1 , ... ,$
the coordinate $\theta$
is globally defined ($2\pi$ periodic)
and so  \me\ is  equivalent to  a free  bosonic string
theory compactified on a circle.\foot{The  trivial
models with $qR=n$  may still look non-trivial  from lower dimensional
Kaluza-Klein point of view.
The equivalence between higher dimensional and lower dimensional descriptions
is,  of course,  established
 once  the contribution of the whole tower of higher massive Kaluza-Klein
states  is taken into account.}
Models with $n <qR< n+1$  are equivalent to  models with  $0 < qR <1$.
This periodicity condition in $qR$  will be  modified in the
superstring
theory: because of the presence of fermions
of half-integer spin   $n$  will be replaced by  $2n$, i.e.   only  models
with $qR=2n$ will be trivial (more generally,  superstring theories with
$(R,q)$  and $(R, q + 2n R\inv)$
will be equivalent).

The Lagrangian  \me\ has  the following useful  form
($x= x_1 + i x_2 = \r e^{i\vp} $) :
\eqn\cov{
L_1 = (\d_a x_i - q \ep _{ij} x_j \d_a y) (\daa x_i - q \ep _{ij} x_j \daa y) +
\da y \daa y  }
$$ = \  D^a x D^*_a x^* + \da y \daa y   \ ,  \ \ \ \  \ \ \ D_a \equiv \da +
i A_a \ , \ \ \   A_a \equiv q\da y \ ,  $$
  where
the  2d gauge potential $A_a$ is flat (locally pure gauge).
 Since $y$ is compact,
the effect of this gauge potential  will be non-trivial if the world sheet will
have non-trivial holonomy (2d  or `world sheet' Aharanov-Bohm effect).

The  flatness of the    potential  $A_a=q \da y$ in \cov\  implies that    $x$
can be
formally `rotated' to  decouple it  from $y$.
Then $y$ satisfies the free-field equation and $x$ is  also expressed in terms
of free fields.
 The only   interaction   which  effectively survives in the final expressions
is the coupling of $x$ to the derivative of the zero mode part of $y$
(e.g.,
$y_*= y_0 + 2\a' p \t + 2Rw\s\ $ in the case of the cylinder as a world sheet).
It is then straightforward to  carry out the canonical quantization procedure,
expressing all observables in terms of free oscillators.
The resulting  Hamiltonian will be  given by the sum
of the free string  Hamiltonian plus  $O(q)$ and $O(q^2)$ terms
depending on the left and right components of the free string angular momentum
operators $\hat J_L$ and $\hat J_R$ \refs{\rut, \rrt}.

 This  bosonic string  model
 is stable in the
non-winding  sector, where there  are  no new  instabilities in
addition to the usual flat space tachyon \rut. This means, in particular,  that
the  Kaluza-Klein field  theory corresponding to   the Melvin  background is
perturbatively stable with respect to the `massless' (graviton, vector,
scalar)  {\it and} massive perturbations (the theory
may still be unstable  at a  non-perturbative  level \gaun).
At the same time,
 there exists a range of parameters $q$ and $ R $
for which  there  are
tachyonic states in the {\it winding }  sector.
This instability  (whose origin is  essentially in
 the gyromagnetic coupling  term
$wq R(\hat J_R -\hat J_L)$ which may have a negative  sign, see below and  cf.
\ene,\ope) is not related to
the presence of the flat
bosonic string tachyon
and survives
also
in the superstring case \rrt.

\subsec{Solution of the superstring  model}

Let us now  consider
the  type II superstring version of \mode\
using first  the   RNS formulation of the model.
The $(1,1)$  world-sheet supersymmetric extension
of the  model \mee,\cov\ has the form ($x^\m\equiv (x^i, y)$)
\eqn\suu { L_{\rm RNS} = G_{\m\n} (x)\del_+  x^\m \del_-  x^\n   +
\l_{R m} {\rm D}_{+} \l^m_R +\l_{L m} {\rm D}_{-} \l^m_L \ , } $$
{\rm D}^m_{\pm n\m} \equiv \delta ^m_n \del_\pm     + \om^m_{\ n \m }\del_\pm
x^\m \ . $$
$\l^m= e^m_\m \l^\m$ are  vierbein components of the 2d
Majorana-Weyl spinors and $\om^m_{\ n \m }$ is  flat   spin   connection (so
that the quartic fermionic terms  are absent).\foot{
In the  basis $e^i= dx^i - q\ep^{ij} x_j dy,  \ e^y= dy$, the spin
connection 1-form  has the following components:
$\om^{ij} =- q\ep^{ij}dy , \ \ \om^{iy} =0.$}
In terms of the left and right   Weyl spinors $\l=\l_1 +i\l_2 $ corresponding
to $x=x_1 +ix_2$ and $\l^y\equiv  \psi$, we get (cf. \cov)\eqn\onn{L_{\rm RNS}=
 \ha ( D_+ x  D^*_- x^* + c.c.) +
\d_+ y \d_- y
  +  \l^*_R  D_+\l_R
 +  \l^*_L D_-  \l_L  }
$$
+ \  \psi_R \d_+ \psi_R +   \psi_L \d_- \psi_L \ , \ \ \ \ \
 D_\pm \equiv  \del_\pm   +  iq \del_\pm  y\ ,     $$
 where the covariant derivative $ D_\pm  $
is the same as in \cov, i.e.  it contains the  flat  2d $U(1)$ potential.
 This means that, as in the bosonic case,
 it is possible to redefine the fields $x$, $\l$
 so that the only non-trivial   coupling  will be to the zero mode of $y$.
Although it may seem that, as in the bosonic case,  the model  with
 $qR=n$
should be  equivalent to the free superstring theory compactified on a circle
(since for $qR=n$
 one can,    in principle,   eliminate the coupling terms in \onn\ by rotating
the fields)
this will not actually be true    unless the integer $n$
 is even,  $n=2k$.
The non-triviality  for $n=2k+1$
is directly related to the presence of space-time
fermions in the spectrum,  which change sign  under $2\pi $ spatial  rotation
accompanying the periodic shift in $y$. This will be obvious in the GS
formulation (see below).

Taking the world sheet to be  a  cylinder $(0<\tau<\infty, 0 < \s \leq \pi) $
so that
 $x,y,\l$   obey the usual
closed-string boundary conditions
\eqn \gam{ x(\t,\s + \pi ) = x(\t,\s)\ , \ \ \ \  \
 y(\t,\s + \pi ) = y(\t,\s) + 2\pi R w \ , \ \ \ \ w = 0, \pm 1 , ... \ ,  }
\eqn\fer{ \l_{R,L} (\t, \s+  \pi) = \pm \l_{R,L} (\t,\s)\ ,  }
 we can solve the classical equations corresponding to
\onn\ by introducing  the    fields  $X$  and $\L_{R,L}$,
which  satisfy the free string equations  but   have
`twisted' boundary conditions ($\s_\pm \equiv \t\pm \s $)\foot{The twist
parameter $\g$  can be interpreted as a flux
corresponding to the 2d  field $A_a = q \del_a y$ on the cylinder,
$\int A= 2qRw\int d\s=2\pi \g$.}
\eqn\bix { x(\t,\s) = e^{-iq y(\t,\s)} X (\t, \s) \ ,   \ \ \ \
\d_+\d_- X = 0 \ , \ \ \ X= X_+ (\s_+) + X_- (\s_-) \ , }
\eqn\bii{  \ \  X(\t,\s + \pi ) = e^{2\pi i \g  }X(\t,\s) \ , \ \
\ \ \ \g \equiv  qRw\ ,  }
\eqn\red{ \l _{R,L}(\t,\s) = e^{-i q y(\t,\s)  } \L_{R,L} (\t,\s)\ ,
\ \ \ \   \  \d_\pm \L _{R,L} =0 \ ,  \ \ \
\L _{R,L} = \L _{R,L} (\s_\mp)  \ , }
\eqn\bou{
 \L _{R,L }(\t,\s + \pi ) = \pm e^{2\pi i  \g   } \L_{R,L} (\t,\s)\ ,   }
 with the signs `$\pm$' in \bou\
corresponding to the Ramond (R)  and Neveu-Schwarz (NS)  sectors.
The crucial observation is that if $x, \l$ are on shell, $y$  still satisfies
the free-field equation:
\eqn\yyyi{ \d_+\d_- y =0 \ , \ \ \ y= y_* + y'\ , \ \ \ y_*= y_0 + 2\a'p\tau +
2Rw\s\ . }
The  explicit expressions for the fields $X=X_+ + X_-$ and $\L_{L,R}$  are then
\eqn\bcfz{
 X_\pm (\s_\pm)  = e^{\pm 2i \g \s_\pm } \X_\pm  (\s_\pm) \ ,  \ \ \ \ \
\X_\pm
(\s_\pm \pm \pi)=
\X_\pm (\s_\pm )\ , }
\eqn\laa{ \L _{L,R}(\s_\pm ) = e^{\pm  2i \g \s_\pm   } \eta_{L,R} (\s_\pm)\ ,
}
where $\X_\pm $  and $\eta_{L,R}$
are the free  fields with the standard free closed string
boundary conditions, i.e. having  the standard oscillator expansions, e.g.,
\eqn\fourie{ \X_+ =  i  \sum_{n \in {\bf Z}} \tilde a_n e^{-2in
\s_+}  \   , \ \
 \ \     \X_- =  i
\sum_{n \in {\bf Z}} a_n  e^{-2in \s_-}  \  , }   $$
 \eta_{R}^{(\rm NS)}=\sum_{r\in {\bf Z}+\ha } c_r\
e^{-2ir\s_- }\  , \  \ \ \ \ \eta_{R}^{(\rm R)} =\sum_{n \in {\bf Z}}
d_n\
e^{-2in\s_-}\ ,
$$
and similar expressions for  the left fermions with oscillators having extra
tildes.
One  can then proceed with the canonical quantization
of the model expressing the observables in terms of the above free oscillators.
It is convenient  to choose   the  light-cone gauge,
eliminating  the oscillator part of $u=y-t$ (see  \refs{\rut,\ruh,\rrt} for
details).
Then the  string states are parametrised by  the following
 global quantum numbers:
the total energy $E$, the Kaluza-Klein linear momentum number $m$
 ($p_y= mR\inv$) and  the winding number $w$,
 the  orbital  momenta $l_R$ and $ l_L $ in the 2-plane
(analogues of the Landau level which replace the
linear momenta $p_1,p_2$) and by  the continuous
momentum $p_3$  (as well as by discrete  momenta
corresponding to  extra 5 toroidal directions).

 The left and right angular momentum operators in the 2-plane  contain the
orbital momentum parts plus the  spin parts
\eqn\eig{  \hat J_{L,R}
 = \pm (l_{L,R} + \ha) + S_{L,R} \   ,  \ \
 \hat J\equiv \hat J_L + \hat J_R = l_L-l_R + S_L + S_R\ ,    }
where  the orbital momenta $l_{L,R}=0,1,2,...  $
and $S_{L,R}$  have  the
standard free superstring expressions \gso\
in terms of free oscillators, e.g., $S_R=
\sum_{n=1}^\infty \big(  b^{\dagger
}_{n+}b_{n+} - b^{\dagger }_{n-}  b_{n-} \big)+$fermionic terms
($b_n$ are bosonic oscillators $a_n$ in \fourie\  rescaled by factors of $(n\pm
\g)^{1/2}$).
In the case when  $\gamma=0$ (or, more
generally,  $\g=n$)
  the zero-mode structure changes in that the translational invariance
in the 2-plane is restored   (the zero mode
oscillators are replaced by  the standard zero-mode operators $x_{1,2},
p_{1,2}$ \refs {\rts , \rut }).
The  number of states operators $\hat N_R$ and $\hat N_L$
have the standard   expressions \gso\  in terms of the oscillators in
\fourie\  so that
$\hat N_{R,L}= N_{R,L} -a ,$
\ $  a^{\rm (R)} =0 , \    a^{\rm (NS) } =\ha $, \ $ N_R= \sum_{n=1}^\infty n
\big( b^{\dagger }_{n+}b_{n+}+ b^{\dagger
}_{n-}b_{n-} + ...) $, etc.
Under the  usual GSO projection
(which is implied by the GS formulation and is necessary for  correspondence
 with the free RNS  superstring theory in the limit $q=0$ but will not
automatically lead to
 space-time  supersymmetry for generic $q$)
  $\N_{R}$  and $\N_{L }$ can    take  only non-negative integer values
They satisfy  the  standard
 constraint $\N_R-  \N_L = mw$.

Computing the stress tensor   one finds
the resulting expression for the  light-cone gauge Hamiltonian\foot{Here
$p_\a^2 $ includes $p_3^2$ as well as the  contributions of  the  linear and
winding  momenta in  other 5
free compactified dimensions (for simplicity
we  shall sometimes set them equal  to zero).}
\eqn\hamil{
\hat \H =\hat \H_0 - \a' q (  Q_L  \hat J_R  + Q_R  { \hat J}_L)    + \ha \a'
q^2 \hat J^2   \ ,   }
$$ \hat \H_0 \equiv
 \ha  \a' \big( -E^2 +  p_\a^2 + \ha Q_L^2 +
\ha Q_R^2  \big) + \hat N_R+  \hat N_L ,
\  \ \ \ \   \  Q_{L,R}= {mR\inv } \pm {
\a'}\inv Rw  . $$
It  can be interpreted as describing   charged states of
 closed superstring compactified on $S^1$
moving  in  the Melvin flux tube background.
$\hat \H$  is different from the  Hamiltonian of the
free string on  a circle $\hat \H_0$  by $O(q)$ (`gyromagnetic' interaction,
cf. \ene,\ope)
 and $O(q^2)$  (charge-independent `gravitational' interaction) terms.
Charged string states are `trapped' by the magnetic field
(they cannot move freely in the 2-plane having discrete
orbital momentum numbers $l_L,l_R$ instead of continuous
linear momenta $p_1,p_2$).

$\hat \H$  can be represented also in the following  (`free superstring
compactified on a circle') form
$$\hat \H =\ha  \a' \big[ -E^2 +  p_\a^2 + (m-qR \hat J)^2 R^{-2}
+  {\a'}^{-2}  w^2R^2 \big]
+ \hat N_R+  \hat N_L -qRw (\hat J_R -\hat J_L) $$
 \eqn\ham{
 =  \ha  \a' \big( -E^2 +  p_\a^2  + {m'}^{2}R^{-2}  +  {\a'}^{-2}  w^2R^2
\big)
+ \hat N_R'+  \hat N_L'
\  ,   }
where\foot{Up to the orbital momentum terms,  $ \hat N_{R,L}'$ can be put into
the same
form  as free operators $ \hat N_{R,L}$
 with  the factor   $n$ replaced by $n \pm  \g$.
This is  related  to  the fact that the model we are solving is
`locally trivial', i.e. the $q$-dependence could be eliminated
by a rotation of coordinates if not for the global effects.}
$ m'\equiv m - qR \hat J $,  \ $  \hat N_R' \equiv  \hat N_R  - \g  \hat J_R$,
\  $   \hat N_L' \equiv  \hat N_L  +  \g  \hat J_L$, \ \  $\g\equiv
qRw$.
The Virasoro  condition $\hat \H=0$  then leads to
the expression for the   mass spectrum $M^2 \equiv E^2 -p_\a^2 $.
The mass spectrum is  invariant under
\eqn\inve{  q\to q+ 2nR\inv\ , \  \ \ \ \ n=0, \pm 1 , ...\ , }
  since  (for $w=0$) this transformation can be compensated by
$m\to m- 2n\hat J=$ integer.  Note that  because  $\hat J$ can take  both
integer
(NS-NS, R-R sectors) {  and}
{\it half-integer}  (NS-R, R-NS sectors) values, the symmetry of the bosonic
part of the spectrum  $q\to q+ nR\inv$  is {\it not}
 a   symmetry of  its
fermionic part, i.e. the full  superstring spectrum is invariant only under
\inve.

The same conclusion about the periodicity in $q$  is  true in  general for
$w\not=0$.    In the form given  above, eq. \hamil\ is  valid  for   $0\leq
w<(qR)\inv $, i.e. for
$0\leq \g < 1$. The generalization to other values of
  $\g $
 is straightforward \refs{ \rut,\rrt}: one is to
   replace $
\g=qRw$  in  \ham\ by  $ \hat \g\equiv \g - [\g]$,
where $[ \g]$ denotes the integer part of $ \g$ ($0\leq \hat\g<1$, cf. \ope).
For fixed radius $R$ the     mass spectrum  is thus periodic in $q$,
  i.e. it is  mapped into itself under \inve\
(combined with $m\to m- 2n\hat J$).
In the case  of $qR =2n$ (i.e.  $\g=2nw=2k$) the
 spectrum is thus equivalent to that of the  free superstring
compactified on a circle.
For $qR= 2n+1$ (i.e.  $\g=(2n +1)w=2k +1$ if $w$ is odd)
the spectrum is the same as that
of  the  free superstring compactified
on a circle with antiperiodic boundary conditions
for space-time fermions \rohm \  (see
also \refs{\att,\kou}).
 This relation will become clear
 in
the GS formulation discussed below.\foot{In particular, it will be apparent
that
the interaction term in the  superstring action
 can be eliminated by a globally defined field transformation
only  if  $qR=2n$, while  for $qR=2n +1$
this can be done at the expense of imposing
antiperiodic boundary conditions (in $\s$ or in the $y$-direction)
 on fermions (under the rotation by
the angle $2\pi qR= 2\pi$  in  the 2-plane,
 which is associated with a periodic shift in $y$,  the bosons remain invariant
but spinors  change sign).}

\subsec{Green-Schwarz formulation }

Given a  generic  curved  bosonic background,  the  corresponding
Green-Schwarz (GS) superstring action  \grs\   defines  a complicated
non-linear  2d theory.
 When one is able to fix a light-cone gauge
and,  moreover, the background geometry is flat
as in the   case of
the Melvin model \mee\   (so that conformal invariance and $\k$-supersymmetry
are guaranteed)
 the  action   becomes  very simple  (cf. \suu)\foot{Its  form
 \refs{\cand,\fra,\barny}
can be explicitly  determined,  e.g.,   by comparing  \fra\ with  the
known light-cone superstring vertex operators \gso.}
\eqn\gss{ L_{GS}
= G_{\m\n} (x)\del_+  x^\m \del_-  x^\n
+  i \S_R {\cal D}_+ \S_R
 +    i \S_L {\cal D}_- \S_L\ , } $$  {\cal D}_a\equiv \del_a  + \four
\g_{mn}\om^{mn}_{ \m }\del_a x^\m
\ .  $$
 Here $S^p_{R,L}$ $\ (p=1,...,8)$ are   the right and left  real spinors of
$SO(8)$ (we consider type IIA theory).
In the  case of \mee\ we get  (cf. \cov)
\eqn\ongs{L_{GS} =   (\del_+ + i q\d_+ y )x (\del_- - i q\d_- y ) x^* +
\d_+ y \d_- y  } $$
  + \   i \S_R (\del_+ - \four q \ep^{ij}\g_{ij}\d_+ y ) \S_R
 +  i \S_L (\del_- - \four q \ep^{ij}\g_{ij}\d_- y ) \S_L
\ .  $$
It is
 natural  to decompose  the $SO(8)$ spinors according
to $SO(8) \to SU(4) \times U(1)$,  i.e.  $\ \S^p_L \to (\S^r_L, \bar \S^r_L)$,
$\ \S^p_R \to (\S^r_R, \bar \S^r_R)$, \ $r=1,..,4$.
 Then the fermionic terms in \ongs\ become
\eqn\jjj{ L_{GS} (\S)= i \bar \S^{r}_R (\del_+ + \ha iq \d_+ y ) \S^r_R +
i \bar \S^{r}_L (\del_- - \ha iq \d_- y ) \S^r_L \ . }
The connection terms in the covariant derivatives in the fermionic part of the
GS action \jjj\ have extra coefficients $\ha$ with respect  to the ones in the
RNS action \onn. This immediately implies that the full   theory is periodic
under $qR \to qR + 2n$.

 The condition that the  GS action \gss,\ongs\ has  residual supersymmetry
invariance  $\S \to \S + \ep(x)$ is  equivalent to
${\cal D}_a  \ep (x(\t,\s))=
\del_a x^\m (\del_\m  + \four  \g_{mn}\om^{mn}_{ \m }) \ep (x) =0$.
The absence of   supersymmetry  invariance is the  consequence of the absence
of  zero modes of  the  above covariant derivative operators,   or,
equivalently,  of the non-existence of  solutions of the  Killing spinor
equation
\eqn\kil{ (\del_\m + \four \g_{mn}\om^{mn} _{\ \ \m}  ) \ep =0 \ . }
 In
the $D=3$ background corresponding to \cov\ $\ep= \ep (x^i,y)$ is  a space-time
 spinor and
$\om^{mn} _{\ \ \m}$ is the same  flat spin connection as in \suu,\onn\ so that
\kil\ reduces  (after $SU(4) \times U(1)$ split of $\ep$ as in \jjj)
to
\eqn\kiii{  (\del_y \mp \ha iq  ) \ep =0 \   . }
The formal solution of \kiii\
$ \ep (y) = \exp (\pm \ha iq )\ \ep (0) $
does not, however, satisfy the
periodic boundary condition in $y$,  $\ \ep (y + 2\pi R)= \ep (y)$
(unless $qR=2n$   when  the Killing spinor does exist,  in agreement with the
fact that in this case the theory is   equivalent to the free
superstring).
The  conclusion is that  for $qR\not=2n$ there is no residual space-time
supersymmetry
in the higher-dimensional (e.g., $D=5$ supergravity)
 counterpart of the $a=\sqrt 3$ Melvin background.\foot{Even though the $D=5$
supergravity background $M^5$ is flat, it is the presence of the flat but
non-trivial spin connection that leads to the breaking of supersymmetry. Let us
note also that
the absence of Killing
spinors in the case of the $a=0$ Melvin solution of the Einstein-Maxwell theory
was  pointed out  in \gib.}

As in the bosonic and RNS cases,   one can   explicitly  solve
the classical
string equations corresponding to \ongs\
(cf. \red,\bou)
\eqn\ttt{\S _{R,L}(\t,\s) = e^{-{i\ov 2}  q y(\t,\s)  } \Sigma_{R,L} (\s_\mp)\
, \ \ \
 \Sigma _{R,L }(\t,\s + \pi ) = e^{ i \pi \g   } \Sigma_{R,L} (\t,\s)\ ,
  }
with the final result that
  the only  essential
difference,  as compared to the free superstring case,
is the coupling of bosons and fermions to the  zero-mode part
of the flat $U(1)$ connection $\del_a y_*$.
The  expressions for the superstring Hamiltonian and mass spectrum are
effectively  the same as
in the RNS approach  \hamil.\foot{For $2k\leq  \gamma<2k+1$
the  operators $\hat N_{L,R}$, $\hat J_{L,R}$  have  the usual  free GS
superstring  form,  which is  similar to their form  in the R-sector of the RNS
formalism with vanishing zero-point energy.
For  $2k-1\leq  \gamma<2k$ the operators  $\hat N_{L,R}$ have
 the `NS-sector' form,  i.e.
they  take half-integer eigenvalues starting from
$-\ha$.}

In general, the  model  with $qR=2n +1$ ($\g= 2k +1$ for odd $w$)
is equivalent to the free superstring compactified on
  a twisted 3-torus (in the limit when the 2-torus part is replaced by
2-plane), or on a circle with antiperiodic boundary conditions for the
fermions \rohm\ (in particular, the theory with $qR=1$ and
$R<\sqrt{2\a'}$ will
have  tachyons).

The fundamental world-sheet fermions  $\S$ that
appear in GS action \gss\ are always {\it periodic}  in $\s$
(this is necessary for supersymmetry of the model in the $q=0$ limit).
This implies that the  `redefined' fermions $\Sigma$ in \ttt\
must   change phase under
a shift in $y$-direction. For  $qR=2n +1$ this results in antiperiodic
boundary conditions for {\it space-time} fermions as functions of $y$
(the space-time fields can be represented, e.g.,    as coefficients in
expansion
of a super string field $\Phi (y,\S, ...)$
in powers of world-sheet fermions).
As a result, there exists a
 1-parameter family of models
interpolating between  the  standard supersymmetric $qR=0$  model with
fermions which are periodic in $y$ and a non-supersymmetric  $qR=1$ model with
fermions which are antiperiodic in $y$.

\subsec{Mass spectrum: supersymmetry breaking and
(in)stability }
The expression for the mass spectrum that follows from
\hamil,\ham\ is
\eqn\kme{
M^2 \equiv E^2 -p_\a^2 =M^2_0 - 2qR\inv m \hat J
-2{\a'}\inv q  R w (\hat J_R -\hat J_L) + q^2 \hat J^2\
}
\eqn\ano{ = 2{\a'}\inv (\N_L+ \N_R)   +  ( m-qR\hat J)^2R^{-2}
 +   {\a'}^{-2}  w^2R^2  - qRw (\hat J_R -\hat J_L)  \ ,  }
where $M_0^2 = 2{\a'}\inv (\N_L+ \N_R)   +   m^2R^{-2}
 +   {\a'}^{-2}  w^2R^2$ is the mass operator of the free superstring
compactified on a
circle.
It is easy to see that  in general $M^2$ is not
positive definite in the winding  ($w\not=0$) sector
because of the last  $O(qRw)$ gyromagnetic interaction term in \ano.
Thus one should expect the presence of instabilities, in agreement
with the magnetic interpretation of the model and the  existence
of charged higher spin states in the spectrum.

Indeed, it  follows from \kme\ that \
(i) the space-time supersymmetry is broken for $qR\not=2n$,\   and  \
(ii) there  exists a range of values of parameters  $q$ and $R$  for which
there are tachyonic states in the  spectrum.

The breaking of supersymmetry is of course expected in view of the magnetic
interpretation of the model
(the coupling is  spin-dependent).  Suppose  that we start with the free
superstring compactified on a circle $y$
 and study
what happens with the spectrum when we switch on the
magnetic field,    $q\not=0$.
Since the mass shift in \kme\ involves  {\it both }
components  $ \hat J_L$ and $ \hat J_R$  of the angular  momentum
 the masses of bosons and fermions that were equal for
$q=0$  will become different for $q\not=0$ (it is impossible to have  both
$\hat J_L$ and $\hat J_R$   equal for
bosons and fermions).
Supersymmetry is absent  already   in the { non-winding}
 sector (where the coupling  is to  the total angular momentum  $\hat J$).

 For
example,
the free superstring  massless  ground states
($\hat N_{L,R}=0=m=w$)  will, according to \kme,    get masses
 $M=|q\hat J|$  proportional to their total angular  momenta,
which   must be integer for bosons and half-integer
 for fermions (cf. \eig).
Note that these
 states are neutral,  so that  from  the
 4-dimensional point of view the shift in the masses
can be interpreted as a gravitational effect.
This  shift implies,
in particular,  that supersymmetry is  broken  already at   the field-theory
($D=5$ or $D=4$  supergravity) level, in agreement
with the absence of Killing spinors in  the  $D=4$
Melvin background  discussed above.

In the absence of supersymmetry  some  instabilities of the bosonic string
model may survive also  in the superstring case.  The
mass operator \ano\ is positive in the non-winding  sector, but, as in the
bosonic case,    tachyonic
states   may   appear  in the winding sector. Consider, for
 example, the  NS-NS    winding states  with zero  Kaluza-Klein
momentum and  zero orbital momentum  quantum numbers    and with
maximal absolute values of  the spins $S_{R,L}$ at given levels
(leading Regge trajectory)
\eqn\sta{ w>0\ , \ \ \  m=0\ , \ \ \ l_R=l_L=0 \ , \ \  \
 S_R=\N_R +1 \ , \ \ \ S_L = -\N_L -1 \ . }
We  shall assume that  $0< qRw <1$ (states with  $w > (qR)\inv $  can be
analysed in a similar way).
Then $\N_R=\N_L\equiv N, $ $ \   \hat J= 0 ,$    \ $    \hat J_R-\hat J_L= 2N +
1 $,  and
\eqn\anot{ \a'M^2 =
4 N  +     {\a'}^{-1}  w^2R^2
 - 2 qR w (2N + 1  ) \ .   }
The  state  with given $N$ and $w$ will be tachyonic  for  $q > q_{ crit},$
$\
q_{crit}=  { 4 N    +   {\a'}^{-1}w^2R^2 \ov 2(2N +1)wR }.$
For $N=0$ we get $\a' q_{crit} = \ha wR$.
  The condition $qRw<1$ is satisfied provided
   $wR < \sqrt {2\a'}$.

 In general,
states with $M^2<0$ can be present only  for $R<\sqrt{2\a' }$, i.e.   the full
spectrum  is {\it tachyon-free}  if $R>\sqrt{2\a'}$.
For fixed $R<\sqrt{2\a' }$  the minimal value of the magnetic field  strength
parameter at which  tachyons  first appear is
$\a' q_{ crit} = \ha R$, corresponding  to the $N=0, w=1$ case discussed above.

All other  sectors (R-R, R-NS, NS-R)  are tachyon-free.
The absence  of  tachyons in the fermionic sectors
is a direct consequence of unitarity.
Since a  unitary tree-level $S$-matrix  should correspond to a
string field theory  with  a hermitian  action,
 the  `square' of hermitian  fermionic kinetic operator should be
positive  in any  background. This translates into  the
positivity of $M^2$ for the
fermionic states in the case of static backgrounds.
One implication is that  similar models
should  have  another  general property of the spectrum:
the  states which become tachyonic  should originate only  from the states
of the free superstring spectrum
which belong to
 the leading Regge trajectory \rrt.
If there were bosonic tachyons not only on the leading  Regge trajectory,
but  also on  the subleading one, then a
fermionic state  with an  `intermediate' value of the spin (but otherwise
the same quantum numbers)
would have $M^2 <0$. Since this is not allowed by unitarity,
in any unitary superstring model corresponding to a static background
 tachyonic states can only appear on the first (bosonic) Regge trajectory.
This is  indeed true in the open superstring case \fer\
and in the  present and more general closed superstring  models  discussed  in
Section 4 \rrt.

Let us note also that the fact that some  higher spin winding
states  may become tachyonic
 means that  that there are new {\it massless}
 states at the critical values of the magnetic field. This
suggests  a possibility of symmetry enhancement
in similar models.  The  magnetic perturbations may  also reveal certain
hidden symmetries of the superstring
spectrum.\foot{Let us note in this connection that a
 special symmetry of a general class of tachyon-free string models
 with finite 1-loop cosmological constant
was discussed in \diens.}

One can consider also the heterotic version of the above model
(where the magnetic field is embedded in the Kaluza-Klein sector)
by combining the `left' or `right' part of the superstring model
with the  free internal part. Then    \rrt\  (cf. \ano)
\eqn\anoh{ \a' M^2 =
2  (\N_R+ \N_L) + p^2_I   +
\a' (mR\inv  -  q  \hat J)^2
 +    {\a'}\inv w^2 R^2
- 2qRw (\hat J_R -\hat J_L)   ,   }
where $ \N_R  -\N_L = mw + \ha p^2_I $,  $ \N_R= 0,1,2, ... ,$  \  $\N_L=
N_L -1 = -1, 0, 1, ...$.
 In addition to  the instabilities  discussed  above
there are  also new ones, which (for the   `self-dual'
value of the  radius $R=\sqrt{\a' }$)  appear  for  infinitesimal values of the
magnetic field. These are the usual Yang-Mills-type  magnetic
instabilities, associated with the   gauge bosons  ($m=w=\pm 1$,
$p_I^2=l_R=l_L=0$, $\hat N_R=N_L=0$, $S_R= 1$, $S_L=0 $)
of the $SU(2)_L$ group.

\subsec{Partition function}
The basic properties of the spectrum are  reflected in the
1-loop (torus)  partition function  $Z$ of the model
which  will be  non-vanishing  for $q\not=2nR\inv$
due to the absence of  the GS
fermionic zero modes, i.e. the absence  of
supersymmetry.
$Z$ is straightforward to compute by computing  the
path integral  in the GS formulation \rrt.
The first step is to  expand $y$ in  eigen-values of the Laplacian on the
2-torus and redefine
the fields $x,x^*$ and $\S_{L,R}, \bar \S_{L,R}$ in \ongs,\jjj\ to eliminate
the non-zero-mode part of $y$ from the   $U(1)$ connection.
The zero-mode part  of $y$ on the torus
($ds^2 = |d \s_1 + \t d \s_2 |^2  , \ \ \t=\t_1 + i\t_2 , \ \
 0<\s_a\leq 1$) is  $y_* =y_0 +
 2\pi R(w \s_1 + w' \s_2)$,  where $w,w'$ are
integer  winding numbers. Integrating over the fields $x,x^*$ and $\S^r_{L,R},
\bar \S^r_{L,R}$,
we get a ratio of determinants of  scalar  operators  of the
type $\del + iA, \ \bd - i\bar A $     with constant connection
$ A=q\del y_* = \pi \chi$, \   $    \chi\equiv    qR (w' -\t w)$.
The partition function has the simple form \rrt\
 \eqn\zzz{  Z(R,  q) = c  V_7 R \int
 {d^2\t \ov  \tau_2^2 }  \sum_{w,w'=-\infty}^{\infty}
 \exp \big( - {\pi (\a' \t_2)\inv R^2 |w' -\t w|^2 } \big) \ } $$ \times \
{\cal
Z}_0 (\t, \bar \t;\chi,\bar \chi )    \  {Y^4 (\t, \bar \t;
\ha \chi ,\ha \bar \chi  ) \ov Y(\t, \bar \t;
 \chi , \bar \chi )}\  ,  $$
where
\eqn\ttyy{
Y (\t,\bar \t; \h, \bar \h) =
 \exp[{{\pi  (\chi-\bar \chi)^2
\ov 2 \t_2}}]  \    \bigg|  {\theta_1(\h| \t)
\ov \h\theta'_1 (0| \t) }  \bigg|^2 \ . }
The factor  ${\cal Z}_0$  in \zzz\  stands for
  the contributions of the integrals over  the constant  fields $x,x^*,
\S_{L,R}, \bar \S_{L,R}$
which become zero modes in the free-theory ($q=0$)
limit
 \eqn\zerr{ {\cal Z}_0 = { (\ha \chi \t_2^{- 1/2})^4 \
(\ha \bar \chi \t_2^{- 1/2})^4 \ov \chi \bar \chi   \t_2^{-1} }=2^{-8} q^6 R^6
 |w' -\t w|^6  \t_2^{-3 } \ . }
$ {\cal Z}_0$    vanishes for  $q\to 0$
in agreement with the restoration of supersymmetry  in this  limit.\foot{The
$q\to 0$ divergence of the bosonic `constant
mode' factor $ \sim  q^{-2}$  corresponds to  the  restoration of the
translational invariance in the $x_1,x_2$-plane  in the zero magnetic field
limit
(this infrared divergence reproduces  the  factor of  the area of the
2-plane).}

The partition function  vanishes at all supersymmetric points $qR=2n$  where
the fermionic
determinants  have   zero modes  ($\theta_1$-functions in $Y$-factors in
\zzz\ have zeros for any $w,w'$). More
generally,  $Z$  is periodic in $q$ (see
\inve)
 \eqn\peri{Z (R,q)= Z(R, q + 2nR^{-1})\ , \ \ \ \ \ n=0, \pm 1, ... \  .}
For $qR= 2n +1$ the partition function   is the same as that of the
 free  superstring compactified
on a circle with antiperiodic boundary conditions
for space-time fermions \rohm\ (the dependence on odd $qR$  can be eliminated
from \jjj\
at the expense of making $\S_{R,L}, \bar \S_{R,L}$
to satisfy antiperiodic boundary conditions in $\s$ or $y$, cf. \ttt).

$Z$ is infrared-divergent for  those  values of the moduli
$q$ and $R$
for which there are tachyonic states in the spectrum
and is finite for  all other values.  In particular, it is finite for $R >
\sqrt {2\a'}$  and arbitrary $q$.  $Z$ is usually interpreted as a cosmological
term or 1-loop effective
potential for the moduli (so that
 one may study its extrema, etc., cf. \giva).
Since \zzz\ is
  non-negative and periodic in $qR$, its minima are at
   supersymmetric points. It is not clear,
 however, that  this interpretation  of $Z$ as an effective potential
applies in the present case
of  a space-time  (i.e. not internal space)  string model
since the $D=4$  background  is curved ($Z$ is rather the full 1-loop effective
action evaluated on a classical solution).

\newsec{Superstring models for  $a=1$ Melvin
and  more general static magnetic flux tube backgrounds }
 In the previous section  we  have discussed the simplest possible
static magnetic flux tube model.
Now we shall consider more general models  \refs{\rut,\rrt} corresponding to
the 2-parameter
flux tube backgrounds \backg.
It turns out that
 the  superstring versions of these models
(which depend on  compactification radius,
vector and axial magnetic field parameters $R, q$ and $\beta$)
have properties analogous to those of the $\b=0$ ($a=\sqrt 3$ Melvin) model.
 In particular, supersymmetry  is  broken for generic values of $(\beta, q$)
and these models
  reduce to the  free  superstring theory  when
both $qR$ and $ \a' \beta  R\inv $ are  even
integers.
The spectrum contains tachyonic states which (in agreement with the general
remarks in Section 2.3)  appear only in the  bosonic NS-NS sector and  belong
to  the leading Regge trajectory.

The  string models  corresponding to \backg\
with $\beta >q $ are
related to the models with $\beta<q$ by the  duality transformation in the
Kaluza-Klein  coordinate $y$.\foot{At the level of the effective action \actt\
this transformation is the usual $T$-duality map,
$\cA \leftrightarrow \pm \cB,   \
\s \to -\s $, $   \ \Phi \to \Phi, $
$\  \hat G_{\m\n}  \to \hat G_{\m\n} , \  \hat H_{\m\n\l} \to \hat H_{\m\n\l}
, $
which is obviously  a symmetry of  \actt.}
More precisely, the $(R, \beta,q)$ model
is  $y$-dual to $(\a'R\inv, q, \beta)$ model so that  the $q=\b$ ($a=1$
  Melvin)    is  the `self-dual' point.
 For fixed $q$ these  models thus fill  the  interval $0 \leq \beta \leq q$
 parametrized by  $\beta$,    with
$a=\sqrt 3$ and $a=1$ Melvin models  being the boundary points.
The non-trivial $\M^3 =(\r,\vp,y)$ part   of the corresponding  bosonic
 string Lagrangian is \rut\  (cf. \mee)
\eqn\lagg{
 L=
    \del_+ \r  \del_- \r
 +   F (\r)  \r^2
(  \del_+ \vp  +   q_+   \del_+ y )
 \  ( \del_- \vp +   q_-  \del_- y )   }
$$
  +  \  \del_+ y  \del_- y  +   {\cal R} [\p_0 +  \ha \ln F (\r) ] \ ,
\ \ \ \ \
  \  F\inv =   1 + \beta ^2  \r^2  \ , \ \ \ q_\pm \equiv q\pm \b \ .  $$
Note that in addition to the metric of $\M^3$ (which is no longer flat)
there is also the antisymmetric tensor and dilaton backgrounds (${\cal R}$ is
2d curvature).
This model is related to   \mee\ by the formal $O(2,2;R)$
duality rotation (combination of a shift of $\vp$ by $y$ and duality in $y$).
Indeed, it can be obtained from  the model which is $y$-dual   to \mee\
 by first changing $q\to \beta, \ \td y \to y$  and then shifting $\vp
\to \vp + q y$. This explains  why this
bosonic model is solvable \refs{\rut,\rrt}  even though the 10-dimensional
target space
geometry is  no longer
flat.
The equivalent form of \lagg\ is
\eqn\lagge{
 L=
    \del_+ \r  \del_- \r
 +   F (\r)
(  \del_+ y -  \beta  \r^2  \del_+ \vp'  )
  (  \del_- y +  \beta  \r^2 \del_- \vp'  )   }
$$
  +\    \r^2  \del_+ \vp'  \del_- \vp'  +   {\cal R} [\p_0 +  \ha \ln F(\r)] \
,
$$
where  we have used the formal notation $\vp'= \vp + q y$. Introducing an
auxiliary 2d vector field with components $V_+, V_-$  we can represent  \lagge\
 as follows, cf. \cov\ (this corresponds to `undoing' the duality
transformation mentioned above)\eqn\lage{
 L=
 \ha ( D_+ x \  D^*_- x^*  + c.c.) +   V_+ V_-  -  V_-   \del_+ y  + V_+
\del_- y   \ , }
$$
D_\pm \equiv \del_\pm  + i\b V_\pm  + iq \del_\pm  y  \  .   $$
Now it is easy to understand why  the classical equations of this model are
explicitly solvable in terms of free fields
and  the partition function is computable.
In spite of the $y$-dependence in the first term, the equation of motion for
$y$  still imposes the constraint that $V_a$ has zero  field strength,
 ${\cal F}(V)=  \del_- V_+ -  \del_+  V_-=0$:
 the variation over $y$ of the first term vanishes once one uses the equation
for $x$ (as follows from the fact that $qy$-terms can be formally absorbed
into a phase of $x$). Then  $V_+ = C_+  +\del_+ \y, \  V_-= { C_-} +  \del_-
\y, \  C_\pm =\const$. In the equations  for $V_+, V_-$ one can again ignore
the variation of the first term in \lage\
since it  vanishes  under ${\cal F}(V)= 0$. We find that
$V_+= C_+ +  \del_+ \y=  \del_+ y, \  V_-= { C_-} +  \del_- \y= - \del_- y$.
The solution of the model then effectively reduces to that  of the model
\mee ,   the only extra non-trivial contribution being the zero-mode parts  of
the two dual fields $y$ and $\y$.
 Interchanging  of $q$ and $\beta$ is  essentially equivalent (after solving
for $C_+,C_-$) to interchanging of  $y$ and $\y$
and thus of  momentum and winding modes.
Eliminating $C_+,C_-$
one gets terms quartic in the angular momentum operators in the final
Hamiltonian.
Similar approach applies to the computation of the partition function  $Z$.
Once $x,x^*$ have been integrated out, the integrals over the constant parts of
$V_+,  V_-$ cannot be easily computed  for $q\beta \not=0$
  and thus   remain in the final expression \refs{\rut.\rrt}.

This  discussion has  a
straightforward generalization to superstring  case.
 The  corresponding RNS action
now contains the  quartic fermionic terms   which reflect
 the  presence of a non-trivial
(generalized) curvature of the space $\M^3$.
The direct analogue of the `first-order' Lagrangian
 \lage\  is  (cf. \onn)\foot{The fermionic part of this Lagrangian is
reminiscent of the fermionic models studied in
\tomb.}
\eqn\lages{
 L_{\rm RNS} =  \ha (  D_+ x \  D^*_- x^*
+ c.c.)  +  \l^*_R  D_+   \l_R  + \l^*_L  D_-
  \l_L }
$$
+\   V_+ V_-  -  V_-   \del_+ y  + V_+   \del_- y   \ .
$$
 The final expressions for the Hamiltonian
and partition function then  look very similar to the bosonic ones
(the role of fermions is just to supersymmetrize the corresponding  free
superstring  number of states and  angular momentum operators  and to cancel
certain normal ordering terms). One finds (cf. \hamil)
\eqn\hail{
\hat \H =
 \ha  \a' ( -E^2 +  p_\a^2  ) + \hat N_R+  \hat N_L }
$$ +\   \ha \a' R^{-2} (m- qR\hat J)^2
+ \ha {\a'}\inv R^2 (w  -  \a' \b R\inv  \hat J)^2
-    \hat \g  (\hat J_R-\hat J_L)\ , $$
where $\N_R-  \N_L = mw$,
\ $  \hat \g\equiv \g - [\g]$, \ $ \g\equiv  qRw + \a' \b R\inv m
- \a' q\b \hat J. \  $
 $[ \g]$ denotes the integer part of $ \g$
and the operators  $\hat N_{R,L}, \ \hat J_{R,L}$ are the same as in \hamil.

 The duality symmetry in the compact Kaluza-Klein direction $y$ (which
interchanges  the axial and vector magnetic field parameters  $\beta $ and $q
$)
is now manifest: \hail\  is
invariant under  $R\leftrightarrow \a'R\inv ,\  \beta \leftrightarrow q,
\ \ m\leftrightarrow w$.
The resulting expression for the mass spectrum
  can  be  written in terms of  the `left'
and `right' magnetic field parameters
and charges, $\  q_\pm \equiv  q\pm \b$, \ \  $Q_{L,R}= {mR\inv } \pm {
\a'}\inv Rw$ (it reduces to  \kme\ when $\b=0$)
\eqn\hamelv{
    M^2   = M_0^2 -   2 (q_+  Q_L \hat J_R + q_- Q_R\hat J_L)
+  \big(q_+^2 \hat J_R+q_-^2 \hat J_L\big) \hat J  \  .}
The only  states which can be tachyonic  are bosonic states  on
the first Regge trajectory with  the maximal value for $S_R$, minimal value for
$S_L$,  and   zero orbital momentum,
i.e.  $\hat J_R=S_R-\ha=\hat N_R+ \ha ,\   $  $\hat J_L=S_L+\ha=-\hat N_L -\ha
$. Then
\eqn\mmm{
\a' M^2   = 2(\hat N_R + \hat N_L)(1 -\hat \g)    + \a' R^{-2} (m- qR\hat J)^2
+  {\a'}\inv R^2 (w  -  \a' \b R\inv  \hat J)^2
   - 2\hat \g  ,  }
which is not positive definite due to the last term $-2\hat \g $.
One finds \rrt\   that for  generic values of $(q,\b)$  there are
instabilities (associated with states with  high spin and charge) for
arbitrarily small values
of the magnetic field parameters.
The special case of $\beta=0$ (or $q=0$),  corresponding to  the $a=\sqrt{3}$
Melvin model discussed in Section 2, is the only exception:
in this  (type II)  model  there are no tachyons
 below some {\it finite} value of $q$.
The  example which illustrates the generic pattern is   the
$a=1$ Melvin model
where   $q=\beta $ ($q_-=0, q_+= 2\b$) and
\eqn\hmelv{
    \a' M^2   =  4 \hat N_R + \a' Q_R^2 -4\hat \g  \hat J_R
\ , \ \ \  \ \  \g =\a' \b  Q_L - \a' \b ^2 \hat J\ .
}
If we choose  for  $R=\sqrt{\a'}$  then  the  states with $w=m$, \    $\hat
N_L=0$,\   $\hat J_ R=\hat N_R+ \ha
$ and
$\hat J_ L= - \ha $
 become tachyonic for $\b$ in the interval
$\beta_{1}< \beta < \beta_{2}$,  \ $
\b_ {1,2}=m^{-1}( 1 \mp  \sqrt {1- \g_{ crit} }\big), \
\g_{crit}={m^2/(m^2+\ha)}. $
For large $m=\hat N_R $ these magnetic field parameters  are very small.
These infinitesimal instabilities appear because
of the presence of states with arbitrarily large charges
and thus are  a special property of  the {\it closed}
 string theory.
Unlike  the usual Yang-Mills-type  magnetic instabilities,   they
(being associated with higher level states)
remain even after the massless-level states get  small masses.
This suggests  that  a  configuration  with generic values of $(q,\b)$
will  `decay' to become a stable  one with special values of $q,\b$.

 The
supersymmetry
is  broken for generic values of the magnetic field parameters
$\beta, q$ (the two  magnetic fields couple to  both $L,R$- components of the
spin  which
cannot  simultaneously  be the same for bosons and fermions).
When $qR=2n_1$  and $\a'\beta R\inv=  2n_2$ , $   \  n_{1,2}=0, \pm
1, ...,  $  the theory is equivalent to the free
superstring compactified on a circle (in this case  $\hat \g =0$ and,
after appropriate shifts of $m,w$ by  integers,  \hail\
reduces to the  free superstring Hamiltonian). If $qR=2n_1+1$
or $\a'\beta R\inv =  2n_2+1 $, then the necessary shift in $m$ or $w$
in the fermionic sector
involves half-integer numbers. In these
cases the theory can be interpreted as a free superstring  on a circle
 with antiperiodic
boundary conditions for space-time fermions.

The corresponding partition function  is
\refs{\rut,\rrt}
 (cf. \zzz)
\eqn\zzzh{  Z(R,  q, \b ) = c  V_7 R \int
 {d^2\t \ov  \tau_2^2 }  \int dC d\bar C  \ ( \a'  \t_2)\inv
 \sum_{w,w'=-\infty}^{\infty}
} $$
\times \exp\big( -  \pi (\a' \b ^2\t_2 )\inv
 [ \h \bar \h   - R (q + \b )  (w'-  \t w) \bar \h
 - R (  q -\b )      (w'- \bar \t w)\h $$
$$
+ \ R^2 q^2   (w'-  \t w)(w'- \bar \t w)]\ \big) \  \times  \ {\cal Z}_0 (\t,
\bar \t;\chi,\bar \chi ) \ {Y^4 (\t, \bar \t;
\ha \chi ,\ha \bar \chi  ) \ov Y(\t, \bar \t;
 \chi , \bar \chi )}
 , $$
where $
  \h   \equiv 2\b C +  qR(w'-\t w)  ,   \
 \bar \chi \equiv  2\b \bar C  +  q R(w'-\bar \t w), $ and
$Y(\t, \bar \t; \chi , \bar \chi )$ and $ {\cal
Z}_0 (\t, \bar \t;\chi,\bar \chi )$ were defined in \ttyy\
and \zerr. The  auxiliary
parameters
  $C,\bar C$ are proportional to the constant parts of  $V_\pm$ in \lages.
 In the limit $\beta \to 0$  we recover the partition function  \zzz\ of  the
model discussed in the previous section.

The   partition function  \zzzh\   has the following  symmetries (cf. \peri)
\eqn\simm{Z(R,q,\beta)=Z(\a'R\inv, \beta, q)  , \ \ \
Z (R,q,\beta)= Z(R, q + 2n_1 R^{-1},\beta +2n_2 {\a'}\inv R) .}
 These are  symmetries of the full conformal field theory (as can be seen
directly  from the string action in the Green-Schwarz formulation).
If  $  qR  \neq n_1 $ and $\a'\beta R^{-1} \neq n_2$, there are tachyons  for
 any value of the radius $R$, and the partition function contains infrared
divergences.
As follows from  \simm,  when  $\a' \beta R\inv $  (or $qR$) is an even
integer,
the partition function reduces to that of the $a=\sqrt{3} $  Melvin
model \zzz .
In particular,  in the special case  when
 both $qR$ and  $\a'\beta R^{-1} $
are even, the partition function is identically zero (then  the theory is
equivalent  to the free superstring). When either $\a' \beta/R$  or $qR$ is an
odd integer, the partition function is finite in a certain range of values of
the radius.

\def \t {\theta}
\def \tt {\td \t}

\newsec{`Twisted' $SU(2) \times U(1)$ WZNW model as
compact analogue of magnetic flux tube  model and
supersymmetry breaking}
The non-trivial part of the 10-dimensional space-time
corresponding to the $a=\sqrt 3$ Melvin model \mee,\onn\
is a flat non-compact space $\M^3$.
The breaking of  supersymmetry  in the  model \onn,\gss\     is  a consequence
of an
incompatibility
between periodicity of space-time spinors in the  Kaluza-Klein direction
$y$
and the presence of
mixing between $y$ and the  angular coordinate of
 2-plane  which  produces a flat but globally non-trivial spin
connection.
Replacing the 2-plane  by a {\it compact}  space  with a non-trivial  isometry
 and mixing   the isometric coordinate with another compact internal
coordinate $y$, one may try to construct
 similar models in which supersymmetry is broken
  while  the  Lorentz
symmetry in the remaining  flat non-compact directions is  preserved.
Below we shall discuss such a model where  the 4-space $R_{x_3}\times
\M^3=(x_3,\r,\vp,y)$
is replaced by $\M^4= SU(2) \ast U(1)$, i.e.  the `twisted' product
of the  $SU(2)$  WZNW model and a circle.
 $\M^4$ (plus  the corresponding torsion)
is locally the  group space $SU(2) \times U(1)$
and thus defines a conformal model.

We shall consider $\M^4$  as (part of) the {\it internal}
 space and
discuss how the twist parameters  lead to  supersymmetry breaking.\foot{A
special case of this  model was  considered
in \kok\ where it  was interpreted as describing
a  space-time magnetic  background (with the curvature of $S^3$ being small,
i.e.  the level of $SU(2)$ being large).}
This will not be in contradiction with the `no-go' theorem
on impossibility of spontaneous supersymmetry breaking by continuous
parameters   \refs{\band,\din}\
since the corresponding (heterotic) string vacuum
will not be supersymmetric already in the absence of the twist
(the central charge condition will not be satisfied
unless one adds  a linear  dilaton background or considers
special values of  the level $k$).
In what follows we shall ignore this
problem, concentrating  on `additional'
supersymmetry breaking induced by  continuous twist parameters.
In view of
a  relation to the magnetic flux tube model,
 this  supersymmetry breaking
can be given a   `magnetic'
interpretation.\foot{In that sense this  model is  similar
 to   other models where supersymmetry is broken (in a discrete way)
by magnetic fields in internal dimensions,
see \bach\ and refs. there.}

Simplest examples of  models  with spontaneous supersymmetry breaking
are  string compactifications
on `twisted' tori (or string analogues of the
`Scherk-Schwarz' \sche\ compactifications)
\refs{\rohm,\sch,\kou}.
 Consider, e.g., the  3-torus
$(x_1,x_2,y)\equiv (x_1 + 2\pi  R' n_1, x_2+ 2\pi  R' n_2 ,y+ 2\pi R m)$
and twist it by imposing the condition that the shift by period in $y$
should be  accompanied by a rotation in the $(x_1,x_2)$-plane.
For a {\it finite } $ R'$ the only possible rotations are by angles
$\ha \pi m$,  i.e.  one may identify the points
$(\theta, y)=(\theta +  2\pi n + \ha \pi  m, y + 2\pi R m), \
\ \cot \theta = x_1/x_2$.
  The superstring theory  with  this  flat but non-trivial
3-space  $\M^3_0$ as   part of the  internal space
was  considered   in \rohm\ (see also \refs{\sch,\kou})
 where it  was found that  such  twist of the torus
breaks  supersymmetry and  leads to  the
existence of tachyons for $R^2<2\a'$
and non-vanishing
 (and finite  for $R^2> 2\a'$)  partition function.
The    $ R'\to \infty$ limit of this  model
is  actually equivalent to the special  case
 $qR=  \four m $  of  the $a=\sqrt 3 $ Melvin    model
\cov\ (the case of  $m=4$  explicitly  considered in \rohm\ is
equivalent to the superstring compactified on a circle with antiperiodic
boundary
conditions
for the fermions).
Since in the  model \cov\  the 2-plane is non-compact and thus the twisting
angle $2\pi qR$ is arbitrary,
 this model   continuously  connects
   large $ R'$ limits of the  models of \rohm\ with different values of
the integer $m$.

Such models with compact  {\it flat}
 internal spaces always have {\it discrete}
 allowed values of the twisting
parameter (a symmetry group of a  lattice   which generates a torus from $R^N$
is discrete). As a result, the supersymmetry breaking  mass scale   $\m$
is directly
proportional to  the compactification mass scale  $R'^{-1} $ \anton.
Given that  $\m$ should be of $TeV$ order,  this implies
the large value for $R'$ and thus the existence
of a tower of `light' ($M\sim  TeV$) Kaluza-Klein states
\refs{\deal,\aaan}. This  leads  to fast growth of coupling with energy
making perturbation theory unapplicable \deal.

It  could happen that  analogous `twistings'
 of models with compact {\it curved} internal spaces with isometries
lead to  vacua
where the supersymmetry breaking scale could
 be continuously adjusted and thus  decoupled from the Kaluza-Klein scale.
Such possibility, however, is   ruled out
 by the results of \refs{\band,\din}.\foot{According to \band\
it is not possible to break supersymmetry  in a continuous way
by adding a marginal perturbation to an $N=2$ supersymmetric  world sheet
conformal model describing  a  space-time
supersymmetric  vacuum of heterotic string theory.
This `no-go' theorem does not apply to the non-compact model
\mee,\onn\
 which can be considered as a marginal
 perturbation  of the
free non-compact model:
the  left-right symmetric
perturbation  $q\ep_{ij}( x_i \del x_j \bd y + x_i \bd x_j \del y) $
in \cov\ { is} marginal  and integrable (since
the  space $\M^3$  is  flat so that $R_{\m\n}=0$
order by order in $q$)  but is not well-defined as a CFT operator.
 Note that according to \hamil,\kme\ the supersymmetry is broken   there
already  at  $O(q)$ order.}
 Even though this will  not resolve the problem of  large
internal dimension (since the `discrete' part  of  the supersymmetry breaking
will already relate the supersymmetry breaking and compactification scales)
it   may  still  be  of interest  to look for other
 continuous  mechanisms of supersymmetry breaking  which  may complement
 the  `discrete' one.  The idea is
 to separate  the issue
of  discrete  supersymmetry breaking due to
non-compensation of the central charge  from the  {\it additional } continuous
supersymmetry breaking induced
by  the twists.


To construct a   generalisation of the model \mode,\mee\
which  will be compact and, at the same time, remain  to be conformal,
let us
 consider
  the $SU(2) \times U(1) $ WZNW  theory
and  `twist' the product  by shifting
the two  isometric Euler angles  $\theta_L$ and $\theta_R$
of $SU(2)$
($ g= {\rm exp}({i\over 2 } \t_L \s_3 )  {\rm exp}({i\over 2 }
\psi \s_2)  {\rm exp} ({i\over 2 } \t_R \s_3  ) $)
 by  the  periodic coordinate  $y\in (0,2\pi R)$
  corresponding to $U(1)$
\eqn\shi{ \theta_L' = \theta_L + q_1 y\ , \ \ \  \ \ \ \theta_R' = \theta_R +
q_2 y\ . }
Here $q_1,q_2$ are
continuous twist parameters.\foot{The    discussion  that follows
can be generalised to the case of
other WZNW models
 (and their  cosets) using the parametrisation
$g = \exp(i\t_L^sH_s) \exp(i\psi^\a E_\a) \exp(i\t_R^s H_s)$ \
(where $H_s$ are  generators of
  the Cartan subalgebra) and mixing $\t_L,\t_R$ with
coordinates of an extra   torus.}
 The special case of such
   model with  $q_2=0$ was  considered  in \kok.
To make contact with the model  \mee\ it  is necessary to keep  both $q_1$
and $q_2$ non-vanishing \rrt.

The resulting  $SU(2) \times U(1) $ WZNW Lagrangian
\eqn\wwww{ L'(q_1,q_2) = L_{SU(2)} (\psi,\theta_L', \theta_R')
+  \del y \bd y  \ ,   }
\eqn\www{ L_{SU(2)} (\psi, \theta_L, \theta_R) =
{k} (\del \psi \bd \psi
+ \del \t_L \bd \t_L + \del \t_R \bd \t_R
+ 2 \cos \psi \del \t_R \bd \t_L ) \ , }
defines a  conformal theory  since
locally   $\M^4= (\psi, \t_L,\t_R,y)$
 is still the same
$SU(2) \times U(1) $ group manifold.
In particular, the corresponding  central charge is unchanged,
$c={ 3k\ov k+2 } + 1$. Its independence of $q_i$
makes it clear that `trivial'  discrete breaking
of supersymmetry due to non-zero central charge deficit
will be unrelated to `non-trivial' continuous  one induced by non-vanishing
`magnetic'
twists $q_i$.

The case of $q_1= -q_2=q$   is  a compact analogue
of the model \mee. Let us  first note that
$L_{SU(2)}$ in  \www\   can be written as
\eqn\ww{ L_{SU(2)} (\psi, \theta, \tt) =
{k}  \big[ \del \psi \bd \psi
+ 4 \sin^2 {\psi \ov 2} \ \del \t \bd \t + 4 \cos^2 {\psi \ov 2}\  \del \tt \bd
\tt  }
$$ + \  2  \cos{\psi } (\del \tt \bd \t - \del \t \bd \tt) \big] \ ,
\ \ \ \ \  \t = \ha (\t_L-\t_R)\ , \   \  \tt= \ha (\t_L+\t_R) \ . $$
For small  $\psi$  (large $k$)   \ww\
reduces to
${k} ( \del \psi \bd \psi
+ \psi^2  \del \t \bd \t + 4 \del \tt \bd \tt +...), $
i.e.  describes  a product
of a 2-disc $(\psi,  \t)$ and a line $ \tt$. Observing that
  for $q_1= -q_2=q$  the shift
\shi\ implies $\t' = \t + q y, \   \tt'=\tt, $
we can establish a  relation between  $L'(q,-q)$ \wwww\ and  \mee\
by identifying the coordinates in the following way:   $ \sqrt k  \psi\to \r,\
\t \to \vp, \  \sqrt k \tt \to x_3$.

The Lagrangian   \wwww\
 can be represented  in the form of  a perturbation of the
$SU(2) \times U(1) $ WZNW model
\eqn\wew{ L' (q_1,q_2) = L_{SU(2)} (\theta_L, \theta_R, \psi)
+ 2q_1 \bar J_3 J_y + 2q_2 J_3 \bar J_y } $$ \  +  \
(1 + kq_1^2 + kq_2^2  + 2k q_1 q_2 \cos \psi )  \del y \bd y \ , $$
where  $J_3= -i k \Tr(\s_3 g\inv \del g),\  \bar J_3 = -i k \Tr(\s_3 \bd g
g\inv ) ,\  J_y,\  \bar J_y $ are the Cartan currents of the $SU(2)\times U(1)$
model\foot{Another way to see why \wwww,\wew\  with $q_1=-q_2=q$
is related to \mee,\cov\ is to use the parametrization
$g =  {\rm exp}({i\over 2\sqrt k  } x_n \s_n ) $
in which (for small $x_n$ or large $k$)
$\ J_3= \sqrt k \del x_3 + \ep_{ij} x_i \del x_j + ... ,
\ \ \bar J_3= \sqrt k \bd x_3 - \ep_{ij} x_i \bd x_j+ ...,   $
$L_{SU(2)}=\del x_i \bd x_i + \del x_3 \bd x_3 + ...,$ $  \ i,j=1,2 $.
Then  $O(q)$ terms in \wew\  coincide with $O(q)$ terms in
$L_1 + \del x_3 \bd x_3$ in \cov.}
\eqn\cur{  J_3 = k(\del \t_L  + \cos \psi \del \t_R) ,
\ \ \
 \bar J_3 = k(\bd  \t_R  + \cos \psi \bd \t_L)  , \ \ \
J_y = \del y  , \ \   \bar J_y = \bd y  .   }
$O(q_1)$ and $O(q_2)$ terms in  \wew\
are thus integrable marginal perturbations
of the $SU(2) \times U(1) $ WZNW model. This follows  directly
from conformal invariance of \wwww\ and is also in agreement with the fact
that  marginal $ J\bar J$-perturbations
by Cartan currents  are integrable \cha.
 Note that it is only in the `chiral' case of  $q_1=0$ or $q_2=0$
  considered in \kok\
that    $L'$ \wwww\  can be represented  as  the original
$SU(2) \times U(1) $  CFT  (with rescaled radius of $y$)
plus   $J\bar J$-term.

The $SU(2) \times U(1) $ group space preserves half of maximal
space-time supersymmetry ($N=4,D=4$); in particular, there
is the corresponding number of Killing spinors.
The  superstring
compactification on $S^3 \times S^1$ was discussed, e.g.,  in \barny\
where it was pointed out that in the context of the effective field theory
approach
the  extra condition  \cand\  on the Killing spinor
$\g^{mnk} \hat H_{mnk}=0$ coming from the  dilatino  transformation law
(under the assumption that  the dilaton is constant)
is not satisfied. This is related to the issue of
cancellation of the central charge, i.e. the presence or absence of the
tree-level potential  term for the  dilaton (`cosmological constant').
For special values of $k$ the central charge can be cancelled
by combining this model with a minimal model  \gep\
or with `untwisted' $N=2$ coset model \kz.
A central charge deficit  can be compensated by
a linear dilaton background with resulting  `discrete'
supersymmetry breaking
\deal.\foot{For a discussion of   supersymmetric $R_Q \times SU(2)_k$-type
  models with linear dilaton   see
\refs{\kounn, \kok}.}
 As mentioned above,  we shall concentrate  on { additional }
supersymmetry  breaking induced by the  continuous
twist parameters $q_i$.

Let us  show that the background associated with  the
twisted model \wwww\  does not admit Killing spinors so that the
corresponding  light-cone gauge
Green-Schwarz  superstring action   does not have residual
supersymmetry.
 In the case of the  WZNW model the fermionic part of  GS  action
is given by \refs{\fra,\barny} (cf. \gss)
\eqn\fef{ L_{GS} (\S) =  i \S_R {\cal D}_+ \S_R
 +    i \S_L {\cal D}_- \S_L\ ,  }
$$
  {\cal D}_\pm \equiv \del_\pm  + \four \g_{mn}
\om^{mn}_{\pm  \m }\del_\pm  x^\m
\ , \ \ \ \  \ \om^{mn}_{\pm  \m } = \om^{mn}_{  \m } \pm \ha  H^{mn}_{\ \ \
\m }     \ . $$
The quartic fermionic terms are absent since the generalised curvature
$R(\omega_\pm)$ vanishes.\foot{The corresponding  supersymmetric \sm
 (RNS) action  was discussed in
 \refs{\hulh,\roh}.  Let us note that the twist \shi\ {\it preserves}
the extended world-sheet supersymmetry the  $\s$-model
action since  the  existence of  the supersymmetry  is determined
by local conditions on a background.
}
If one chooses the left-invariant  vierbein basis ($e^m= e^m_\m dx^\m =- i \Tr
(\s^m g\inv dg)$)
 then one
of the two generalised connections vanishes ($H_{\m\n\l} = -f_{mnk} e^m_\m
e^n_\n e^k_\l$):
$\ \om^{mn}_{-\m }=0, \ \om^{mn}_{+\m }= - {3\ov 2} f^{mn}_{\ \ \m}$ (in the
right-invariant basis $\om^{mn}_{+\m }=0$).
As a result,  half of the maximal supersymmetry is preserved
since $\S_R$ fermions remain free
 (the corresponding Killing spinor  equation \kil\ with $\om \to \om_+$ has
$\epsilon$=const  as a solution).
It may seem that the same conclusion should be true also
in the `shifted' theory \wwww\
since  locally \shi\ can be considered as a coordinate transformation
and thus  the transformed $\om^{mn}_{-\m }$  should also vanish.
However,
the left-invariant vierbein  basis $e^m_\m$
 which    is independent of $\t_L$
 explicitly depends on $ \cos \t_R, \ \sin \t_R$ and thus its direct analogue
obtained by making  the shift \shi\ is  not  defined
 unless $q_iR$ are integers.
 If one  uses
 the original left-invariant basis,  one  needs to make
 an extra  local Lorentz transformation
to make   the full metric diagonal. Then
${\cal D}_-$  gets a  flat   but
non-trivial $q_i$-dependent connection term.
As a result,  for generic $q_i R\not= 2n_i$ the Killing spinor equation ${\cal
D}_{-\m} \ep=0$
 has  no  solutions consistent with
periodic boundary conditions in $y$ (cf. \ongs,\jjj,\kiii), i.e.
supersymmetry is completely broken.

This can be seen more explicitly  by  choosing
 another  (`isometry-adapted')
 basis  corresponding to the diagonal form of the WZNW action \ww: $e^1=d\psi,
\ e^2= 2 \sin {\psi\ov 2} d\t, \ e^3= 2 \cos{\psi\ov 2} d\tt, \ e^y =dy$.
Then $\om^{\t\psi}=-\cos{\psi\ov 2}d \t, \ \om^{\tt\psi}=\sin{\psi\ov 2}d \t,
\ \om^{\t\tt}=0, \  H= \ha e^\psi \wedge e^\t \wedge e^{\tt}$,  \
$\om^{\t\psi}_\pm =-\cos{\psi\ov 2}d( \t \pm \tt),
\  \om^{\tt\psi}_\pm =\sin {\psi\ov 2}d( \tt \pm \t)$.
The  solution of the corresponding Killing equations  is
$ \ep_{R,L} (\psi, \t, \tt)   =
  {\rm exp}( {i\over 2 } \psi \s_3 )  {\rm exp}( \pm {i\over 2 }
\t_{R,L} \s_1) \ep_{R,L} (0),    $
where $\t_{L,R}=\tt \pm \t$ and $\ep_{L,R} (0)$=const.
Making the  transformation \shi\  we find that the formal solutions of the
Killing spinor equations corresponding to the `twisted' model are  (cf.
\jjj,\kiii):
\eqn\killl{ \ep_{L}  = { e}^{{i\over 2 } \psi \s_3 }
 { e}^{{-{i\over 2 } \t_{L} \s_1}}
{ e}^{{-{i\over 2 } q_1 y \s_1}}
 \ep_{L} (0)\ , \ \ \ \   \ep_{R}  = { e}^{{i\over 2 } \psi \s_3 }
 { e}^{{{i\over 2 } \t_{R} \s_1}}
{ e}^{{{i\over 2 } q_2 y \s_1}}
 \ep_{L} (0)\ .}
As in \kiii\ the periodic boundary conditions in $y$ imply that
 supersymmetry is broken unless $q_{i} R = 2n_{i}$.

The dependence of the Killing spinors on $\t_L$ and $\t_R$
is related to the existence of the fixed points ($\psi=0,2\pi$)
in the action of the
isometries corresponding to shifts along $\t_L$ and $\t_R$.
The same is true in the 2-plane  case in the polar coordinate basis.
 The  breaking of supersymmetry by the twist \shi\ may  be attributed to this
dependence. Similar breaking will
  thus happen in general when an isometry which has fixed points   is `mixed'
with another circular dimension.
This dependence of Killing spinors on angular coordinates
 leads  also to  an   apparent  `breakdown' of supersymmetry \kal\  after the
duality transformations
in these coordinates  (it is only the  local realisation  of supersymmetry
that is  actually broken by the duality \refs{\bak,\ssu}).
 Note that the shift \shi\ becomes part of the $O(3,3;Z)$
duality transformation group  of the $SU(2) \times U(1)$  model \hass\ only
when $q_iR=2n_i $, i.e. when supersymmetry is unbroken.

It would be interesting to determine the spectrum of the model \wwww\
to see the supersymmetry breaking explicitly. This can probably be  done
by  generalizing the approach of \kok\   where the special case
 of $q_2=0$  was solved. It is clear from the
 spectrum given in \kok\   that  $q_1$ plays the role of the the supersymmetry
breaking parameter.

To  try to construct `realistic'
 models  which include  this `magnetic'  supersymmetry breaking  one needs
 to address the question
of  saturation  of the central charge condition.
One possible suggestion  is to relax this condition,
assuming that the dilaton equation
should eventually be satisfied with loop and non-perturbative corrections
included \refs{\deal, \bach} (note, however,  that  $\delta c \sim 1/k$ is
small only if $k$ is large and that
 leads back to the problem of large compactification scale).
Another is to consider special values of $k$ for which
the total $c$ can be balanced by combining this  model, e.g.,
with $N=2$ coset one.
An interesting aspect of  the supersymmetry breaking  induced
by the  `magnetic'
twists is that the resulting contribution to the   cosmological constant
is likely to be very small.
Assuming that  the compactification radius is $R\sim \sqrt { \a'}
 \sim M\inv_{Pl}$
 while the `magnetic'
supersymmetry breaking scale $\m^2 \sim
q$ is  of order of $TeV^2$ or $M^2_W$,
and  that  the partition function of a `realistic'  model
 will be similar to \zzz\  (which  is proportional to $q^6$ for small $q$),
 one may
expect  to find
$\Lambda_4 \sim M^{12}_W/M^8_{Pl}$ which is very small indeed.

\vskip 1cm
 \noindent {\bf Acknowledgements}

\noindent
I would like to thank  T. Banks,  M. Green,  E. Kiritsis,
 K. Kounnas, D. L\"ust, F. Quevedo  and M. Tsypin
for  useful discussions and remarks. I am grateful to
  J. Russo   for  collaboration and  discussions.
I  acknowledge also the
support
of PPARC and of  ECC grant SC1$^*$-CT92-0789.

\vfill\eject
  \listrefs
\end

\\
Title:  Closed superstrings in magnetic field
Author: A.A. Tseytlin
Comments:  28 pages,  harvmac
(extended version: contains a new section discussing a=1 Melvin
and more general static magnetic flux tube models)
Report-no: CERN-TH/95-215,  Imperial/TP/94-95/57
\\

\\
\newsec{Conclusions }
The  simple model considered in the main part of  this paper  describes
  type II superstring  moving in  a flat  but topologically non-trivial
10-dimensional space.
The non-trivial  3-dimensional part of this space   is a `twisted' product
 of a 2-plane and a circle $S^1$  (the periodic shifts in the   coordinate  of
$S^1$
being  accompanied  by rotations in the plane). The free continuous
moduli parameters are the radius  $R$ of $S^1$  and the `twist' $q$.
If    other 5 spatial
dimensions are toroidally compactified,  the model can be interpreted as
corresponding
 to the Kaluza-Klein Melvin magnetic flux tube background in 4 dimensions
($R$ being Kaluza-Klein radius and $q$ being proportional to the magnetic
field strength).

This   model  can  be  easily solved either
in the RNS or light-cone GS approach and
exhibits several  interesting features.
 The supersymmetry is broken if $qR\not=2n$.
For $qR=2n$ the theory is equivalent to the standard free superstring theory
compactified on a circle with periodic boundary conditions for space-time
fermions;   for $qR=2n+1$ it is equivalent to the free superstring with
antiperiodic boundary conditions for the  fermions (the model thus continuously
interpolates between these two free superstring models).
more general static magnetic flux tube
 models of \rut\ (which depend on  compactification radius,
vector and axial magnetic field parameters $R, q$ and $\beta$)
have analogous properties.
 In particular, supersymmetry  is  broken for all of these
models (for generic values of $\beta, q$).
These more general
models  reduce to the  free  superstring theory  when
both $qR$ and $ \a' \beta  R\inv $ are  even
integers.